\documentclass[12pt]{iopart}
\usepackage{amsfonts}  
\usepackage{color}

\begin{document} 

\title[VCS representations, induced representations, and geometric
quantization I]{Vector coherent state representations, induced
  representations, and geometric quantization: \\
  I.\ Scalar coherent state representations}

\author{S D Bartlett\dag\ddag, D J Rowe\dag\ and J Repka\S}
\address{\dag\ Department of Physics, University of Toronto, 
  Toronto, Ontario M5S 1A7, Canada}
\address{\ddag\ Department of Physics, Macquarie University, 
  Sydney, New South Wales 2109, Australia}
\address{\S\ Department of Mathematics, University of Toronto, 
  Toronto, Ontario M5S 3G3, Canada}

\ead{stephen.bartlett@mq.edu.au}

\begin{abstract}
  Coherent state theory is shown to reproduce three categories of
  representations of the spectrum generating algebra for an algebraic
  model: (i) classical realizations which are the starting point for
  geometric quantization; (ii) induced unitary representations
  corresponding to prequantization; and (iii) irreducible unitary
  representations obtained in geometric quantization by choice of a
  polarization.  These representations establish an intimate relation
  between coherent state theory and geometric quantization in the
  context of induced representations.
\end{abstract}

\submitto{\JPA}

\section{Introduction} 

The process of quantizing a classical system has been of interest
since the early days of quantum mechanics and remains an active field
of research.  The most sophisticated quantization procedure is
provided by \emph{geometric quantization} (GQ).  This procedure,
founded on Kirillov's so-called {\it orbit method} \cite{K1}, was
inititated by Kostant~\cite{kostant}, and Souriau~\cite{souriau}, who
extended the orbit method by capitalizing on the physical insights
that come from its applications to quantum mechanics. Thus geometric
quantization incorporates ideas that physicists have used for many
years, but provides new avenues to address quantization of more
difficult systems.

Our primary concern in this paper is the significance of GQ for the
representation theory of Lie groups and Lie algebras. Its relevance to
representation theory is based on the observation that quantizing a
model with an SGA (spectrum generating algebra), defined in the
following section, is equivalent to constructing an appropriate
irreducible unitary representation of that algebra~\cite{thesis}.
Conversely, as brought to light by GQ, the construction of a unitary
irrep of a Lie group or algebra is often equivalent to quantizing some
classical Hamiltonian system.  Thus, the theory of induced
representations plays a central role in the quantization of a model
and in quantum mechanics in general, as emphasized by
Mackey~\cite{mackey}.  Establishing an explicit correspondence between
induced representation theory and GQ sheds new light on both theories.

The theory of induced representations is viewed in this paper from the
perspective of coherent state theory~\cite{per86,Klauder,Onofri,RR91},
which incorporates other inducing constructions in terms of structures
and concepts that relate naturally to quantum mechanics and GQ.  It is
shown that coherent state theory reproduces three categories of
representations of the SGA for an algebraic model, and that these
categories are related to structures in geometric quantization.
First, there are classical realizations of the SGA as functions on a
phase space; GQ begins with such a classical realization.  The
coherent state construction also yields the induced unitary
(reducible) representations that correspond to prequantization.
Finally, the unitary irreducible representations, corresponding to
full quantization, are obtained through means related to the choice of
a polarization in GQ.  These techniques are illustrated by a variety
of examples.

A relationship between cohererent state theory and geometric
quantization follows naturally from the problem that gave rise to
coherent state representations; namely, ``construct an irrreducible
unitary representation of a group from the properties of a set of
coherent states''~\cite{Onofri,OnofriPauri}.  As will be shown in the
following, the expectation values of the Lie algebra over a set of
coherent states give a classical representation of the Lie algebra.
Thus, regaining a unitary irrep is equivalent to quantization of this
classical representation.  It is found that reconstructing the irrep
is possible in coherent state theory only for particular orbits which
give rise to the classical representations that are described as {\em
  quantizable} in geometric quantization.

It is the principal aim of this paper to show that coherent state
methods provide an intuitive and practical framework for implementing
the procedures of induced representations and geometric quantization.
We thereby attempt to make these fundamental theories accessible to a
wider community. The illustrative examples presented here are
intentionally simple.  However, it is noted that all three theories
have been deployed in non-trivial ways. The theory of induced
representations has made seminal contributions in physics, e.g., to
the representation theory of the Poincar\'e group \cite{Wig39} and the
space groups of crystals \cite{Zak}; indeed, it is central to
representation theory and quantum mechanics \cite{mackey}.  Geometric
quantization leads to a deep understanding of the route from classical
mechanics to quantum mechanics.  It has been used, for example, to
develop theories of quantized vortices in hydrodynamics \cite{diffeo}
and nuclei \cite{RI79} and to reproduce the irreducible unitary
representations of compact Lie groups and the holomorphic discrete
series of reductive Lie groups \cite{zhao}. Other applications can be
found in the book of Guillemin and Sternberg \cite{GS}.  Similarly,
coherent state theory has provided a fundamental understanding of
classical-like behaviour in quantum mechanics, e.g., in the field of
quantum optics \cite{Glauber}. It has also been used
extensively~\cite{Klauder}, particularly in its vector coherent state
extension~\cite{vcs} (outlined in the following paper), in the
quantization of numerous physical models by explicit construction of
the irreducible representations of their spectrum generating algebras
(cf., ref.~\cite{RR91} for a long list of references).  Our hope is
that the complementary aspects of these powerful methods will result
in their successful application to even more challenging problems.

It has long been known \cite{Rawnsley} that there is a close
relationship between the theories of coherent states, induced
representations and geometric quantization.  Indeed, the Kirillov
method \cite{K1,kirillov,Kir99}, from which geometric quantization
emerged, was expressed as an inducing construction.  This relationship
was used, for example, by Streater \cite{streater} to construct the
irreducible unitary representations of the semidirect product
oscillator group, following both Mackey and Kirillov methods, and by
Dunne \cite{dunne} to construct the irreps of $SO(2,1)$.  It is
particularly well known that the coadjoints orbits of a Lie group play
central roles in all three theories (cf., for example, Chapter 15 of
Kirillov's book \cite{kirillov}). The contribution of this paper is to
explore the detailed relationships between the theories so that their
complementary strengths may be better understood and more readily
applied to problems in physics.  Moreover, by establishing detailed
relationships, we hope to make it easier to move between the different
expressions of quantization and induced representations in the
solution of complex problems.

With a relationship between scalar coherent state theory and GQ in
place, the generalization to vector coherent state
theory~\cite{vcs,RR91}, in which irreps of an algebra are induced from
multidimensional irreps of a subalgebra, indicates new ways of
applying geometric quantization to models with intrinsic degrees of
freedom.  Conversely, the different perspective of geometric
quantization suggests possibilities for generalization of the theory
of induced representations.  These issues will be discussed in the
sequel paper.

\section{Algebraic models}

An algebraic model is defined below as a model with an SGA.  The
quantization of an algebraic model and the construction of the
irreducible unitary representations of its SGA are then related
problems.  However, whereas quantization starts with the classical
Hamiltonian dynamics on a phase space, the theory of induced
representations starts with the abstract SGA.  These and other
concepts invoked in the two constructions are reviewed in this
section.

\subsection{Observables and spectrum generating algebras} 

In classical mechanics, observables are realized as smooth
real-valued functions on a connected phase space $\mathcal{M}$, i.e.,
elements of $C^\infty(\mathcal{M})$.  They form an
infinite-dimensional Lie algebra with Lie product given by a Poisson
bracket.  In quantum mechanics, observables are interpreted as
Hermitian linear operators on a Hilbert space $\mathbb{H}$; they are
elements of $GL(\mathbb{H})$ and form an infinite-dimensional Lie
algebra with Lie product given by commutation.

The algebras $C^\infty(\mathcal{M})$ and $GL(\mathbb{H})$ for a given
physical system are different \cite{joseph}.  However, a simple
relationship may be established between finite-dimensional
subalgebras of $C^\infty(\mathcal{M})$ and $GL(\mathbb{H})$.  Thus, it
is convenient to consider an abstract Lie algebra of observables
$\mathfrak{g}$ that is real and finite-dimensional and can be
represented classically by a homomorphism $J:\mathfrak{g}\to
C^\infty(\mathcal{M})$ and quantum mechanically by a unitary
representation $T:\mathfrak{g} \to GL(\mathbb{H})$.  Let $\mathcal{A}
= J(A)$ and $\hat A = T(A)$ denote classical and quantal
representations, respectively, of an element $A\in\mathfrak{g}$.

Then, if elements $A$, $B$, and $C\in \mathfrak{g}$ satisfy the
commutation relations
\begin{equation}
  [A,B] = \rmi \hbar \, C \, ,
\end{equation}
the corresponding linear operators and functions satisfy
\begin{equation}
  [\hat A,\hat B] = \rmi \hbar \, \hat C \, ,
\end{equation}
and
\begin{equation}
  \{ \mathcal{A} , \mathcal{B} \} = \mathcal{C} \, ,
\end{equation}
where $\{ \; ,\;\}$ denotes the classical Poisson
bracket.\footnote{Note that we follow the practice, common in quantum
  mechanics, of representing the infinitesimal generators of a unitary
  group representation, which correspond to physical observables, by
  Hermitian operators.  To regard the real linear span of such
  operators as a real Lie algebra then requires inclusion of a factor
  $\rmi$ in the commutation relations. The Poisson bracket needs no such
  factor.} (More precisely, the classical homomorphism is given by $A
\to \rmi\hbar \mathcal{A}$, so that $\{ (\rmi\hbar\mathcal{A}) ,
(\rmi\hbar\mathcal{B}) \} = \rmi\hbar (\rmi\hbar\mathcal{C})$.)

Let $\mathcal{G}\subset C^\infty(\mathcal{M})$ denote the classical
algebra $\mathcal{G} =\{ J(A) | A\in \mathfrak{g}\}$.  We shall
describe the algebra $\mathcal{G}$ (and hence the algebra
$\mathfrak{g}$) as a \emph{spectrum generating algebra} for the
classical system if the values of the observables in $\mathcal{G}$ are
sufficient to uniquely specify a point in $\mathcal{M}$ and their
gradients span the tangent space of $\mathcal{M}$ at every point.
(Other compatible definitions of a SGA can be found in the
literature~\cite{ihrig}.  For example, a
classical SGA can be defined by requiring that the only functions in
$C^\infty(\mathcal{M})$ that Poisson commute with all elements of
$\mathcal{G}$ are the constant functions~\cite{zhao}.)

A finite-dimensional Lie algebra $\mathfrak{g}$ is said to be a SGA
for a quantal system if the Hilbert space for the system carries an
irreducible representation of $\mathfrak{g}$.  However, to be useful,
one may require also that the Hamiltonian and other important
observables of the system should be simply expressible in terms of
$\mathfrak{g}$, e.g., by belonging to its universal enveloping
algebra.  Thus, in quantizing a classical model, we shall seek a
quantal system with the same SGA as the classical model.  A model
dynamical system that has a finite-dimensional SGA is said to be an
\emph{algebraic model}.

\subsection{Phase spaces as coadjoint orbits} 
\label{sect:coadorbits}

Let $G$ be a group of canonical transformations (i.e.,
symplectomorphisms) of a classical phase space $\mathcal{M}$ for a
model.  Then, if $G$ acts transitively on $\mathcal{M}$, it is said to
be a {\em dynamical group} for the model.  If an element $g\in G$
sends a point $m\in\mathcal{M}$ to $m\cdot g\in\mathcal{M}$, then
$\mathcal{M}$ is the group orbit
\begin{equation}
  \mathcal{M} = \{ m\cdot g\, |\, g\in G\} 
\end{equation}
and diffeomorphic to the factor space $H_m\backslash G$
with isotropy subgroup
\begin{equation}
  \label{eq:PhaseSpaceIsotropySubgroup}
  H_m = \{ h\in g \,|\, m\cdot h = m\} \, .
\end{equation}
A remarkable fact~\cite{kostant,souriau} that will be used extensively
in the following is that a phase space with a dynamical group can be
identified with a coadjoint orbit.  Conversely, every coadjoint orbit
is a phase space.  Moreover, the Lie algebra $\mathfrak{g}$ of $G$ is
a SGA for the model.  Note that in quantum mechanics, one is often
interested in projective representations of a given dynamical group.
Thus, we consider projective as well as true representations; in
practice, it is often simpler to choose a dynamical group that is
simply connected so that all its representations are true
representations.

Recall~\cite{marsden} that $G$ has a natural adjoint action on its Lie
algebra $\mathfrak{g}$
\begin{equation}
  {\rm Ad}(g) : \mathfrak{g} \to \mathfrak{g} ;\ A \mapsto A(g)=
  {\rm Ad}(g) A \, ,
\end{equation}
where, for a matrix group, ${\rm Ad}(g) A = gAg^{-1}$.  $G$ also has a
coadjoint action on the space $\mathfrak{g}^*$ of real-valued linear
functionals on $\mathfrak{g}$ (the dual of $\mathfrak{g}$).  The
action of an element $\rho\in\mathfrak{g}^*$ on an element
$A\in\mathfrak{g}$ is given by the natural pairing of $\mathfrak{g}$
and $\mathfrak{g}^*$, expressed as $\rho(A)={\rm Tr}(\rho A)$ for
matrices.  The coadjoint action is then defined, for $\rho\in
\mathfrak{g}^*$ and $g\in G$, by $\rho\to\rho_g$, where
\begin{equation}
   \rho_g(A) = \rho(A(g)) \,,\quad \forall\, A\in \mathfrak{g}\,.
\end{equation}
Thus, the coadjoint orbit
\begin{equation}
  \mathcal{O}_\rho = \{ \rho_{g} \, |\, g\in G \} \, , 
\end{equation}
is diffeomorphic to the factor space $H_\rho\backslash G$ with
isotropy subgroup $H_\rho = \{ h\in g \,|\, \rho_h = \rho\}$.  We
shall refer to an element $\rho$ of $\mathfrak{g}^*$ as a
\emph{density}.  Now if a density $\rho\in\mathfrak{g}^*$ is chosen
such that $H_\rho=H_m$ then there is a diffeomorphism $\mathcal{M}\to
\mathcal{O}_\rho$ in which $m \mapsto \rho$ and $m\cdot g \mapsto
\rho_g$. This map is known as a {\em moment
  map}~\cite{kostant,souriau,marsden}.

Such a moment map defines a classical representation
$J:\mathfrak{g}\to\mathcal{G}; A\mapsto \mathcal{A}= J(A)$ of the Lie
algebra $\mathfrak{g}$ as functions over the classical phase space
$\mathcal{M}$, defined by
\begin{equation}
 \mathcal{A}(m\cdot g) = \rho_g (A)\,,
\end{equation}
with Poisson bracket given by
\begin{equation}
  \label{eq:PoissonBracket}
  \{\mathcal{A},\mathcal{B}\} = \omega (A,B) \, ,
\end{equation}
where $\omega$ is the antisymmetric two-form\footnote{With the
  realization of the Lie algebra $\mathfrak{g}$ as a set of invariant
  vector fields on $\mathcal{M}$, $\omega$ becomes a two-form on
  $\mathcal{M}$.} with values at $m\cdot g\in \mathcal{M}$ given by
\begin{equation}
  \label{eq:OmegaAsDRho}
  \omega_{m\cdot g} (A,B) = - \frac{\rmi}{\hbar} \, \rho_g ([A,B])
  \,,\quad \forall\, A,B\in \mathfrak{g}\,.
\end{equation}
Thus, the moment map $\mathcal{M} \to \mathcal{O}_\rho$ defines a
symplectic form $\omega$ on $\mathcal{O}_\rho$.  This form is known to
be nondegenerate and the map $\mathcal{M}\to\mathcal{O}_m$ is a
symplectomorphism.

\subsection{Geometric quantization}

Geometric quantization formalizes the ideas of Dirac~\cite{diractext}
concerning the quantization of a classical system.  The idea is to
replace the classical (Poisson bracket) algebra of functions on a
phase space by a unitary representation; i.e., map each classical
observable (function) $\mathcal{F}$ to a Hermitian operator $\hat{F}$
such that, if
\begin{equation}
  \label{eq:RulesOfQuantization1}
  \{\mathcal{F}_1 , \mathcal{F}_2 \} = \mathcal{F}_3 \, ,
\end{equation}
then
\begin{equation}
  \label{eq:RulesOfQuantization2}
  [\hat{F}_1,\hat{F}_2] = \rmi\hbar \, \hat{F}_3 \, .
\end{equation}
There are some additional requirements of the \emph{Dirac map.}
First, all constant functions must map to multiples of the identity
operator $\hat{I}$.  Second, the unitary representation should be
irreducible.

GQ solves the first part of the Dirac problem by a construction known
as prequantization.  However, it is now known that there are generally
no irreducible unitary representations of the full algebra of
observables~\cite{joseph}.  Thus, a complete solution of the Dirac
problem is impossible.  When there exists a finite-dimensional SGA for
the model then, under certain quantizability conditions, the
restriction to the SGA of the representation given by prequantization
becomes fully reducible.  Moreover, GQ gives a prescription for the
reduction in terms of a \emph{polarization}.  For details of the
techniques of GQ, see Woodhouse~\cite{woodhouse}.

It will be shown in the following that the reducible representation of
a SGA given by prequantization of a classical phase space
diffeomorphic to $H_\rho \backslash G$ integrates to the reducible
representation of $G$ induced from a one-dimensional representation of
$H_\rho$ in the standard theory of induced representations.  Likewise,
the irreducible representations obtained by introducing a polarization
are obtainable by the coherent state inducing construction.

\section{Scalar coherent state representations}\label{sec:Scalar}

In this section, we show how the coherent state construction
reproduces the three categories of representations of the SGA of an
algebraic model: classical realizations, reducible unitary
representations corresponding to prequantization, and irreducible
unitary representations of a full quantization.

Let $G$ with Lie algebra $\mathfrak{g}$ denote the dynamical group of
an algebraic model and let $T$ denote an abstract (possibly
projective) unitary representation of $G$ on a Hilbert space
$\mathbb{H}$.  There is no need to make a precise specification of $T$. 
For example, if $G$ has a right invariant measure $\rmd v(g)$,
$\mathbb{H}$ could be the space
$\mathcal{L}^2(G)$ of square integrable functions with respect to this
measure and $T$ the regular representation.  Or, for application to
models of many-particle systems, $\mathbb{H}$ might be the standard
many-particle Hilbert space $\mathcal{L}^2(\mathbb{R}^{3N})$ of
square-integrable functions of many-particle Cartesian coordinates
and $T$ a Weil~\cite{kashiwara} or Schr\"odinger representation (see
section~\ref{subsubsec:Schrodinger}).

For notational convenience we denote the representation $T(A)$ of an
element $A\in \mathfrak{g}$ by $\hat A\equiv T(A)$.

\subsection{Classical realizations of $\mathfrak{g}$}
\label{sect:classical rep}

In section~\ref{sect:coadorbits}, the classical phase space for an
algebraic model is shown to be expressible as a coadjoint orbit.  
In this section, we show that these coadjoint
orbits can be projected from group orbits in the Hilbert space.

For any state $|0\rangle\in\mathbb{H}$ of unit norm (i.e., $\langle
0|0\rangle =1$) there is a system of coherent
states~\cite{per86}
\begin{equation}
  \{ |g\rangle = T(g^{-1})|0\rangle ; g\in G\} 
\end{equation}
and a corresponding set of dual states
\begin{equation}
     \{ \langle g| = \langle 0|T(g) ; g\in G\} \,.
\end{equation}
Moreover, there is a natural identification of any state
with a density, i.e., an element of the dual algebra $\mathfrak{g}^*$.
Thus, the coherent states define a system of densities
\begin{eqnarray}
  \rho(A) = \langle 0|\hat A|0\rangle \, ,   \\
  \rho_g(A) = \langle g|\hat A |g\rangle = \langle 0| \hat A(g) |0\rangle
  = \rho(A(g)) \, , \quad g\in G\,,
\end{eqnarray}
where
\begin{equation} \label{eq:17}
  \hat A(g) =T( {\rm Ad} (g)A) = T(g)\, \hat A \, T(g^{-1}) \, ,
  \quad A\in \mathfrak{g} \, .
\end{equation}
It follows that the coherent states determine a coadjoint orbit
\begin{equation}
  \label{eq:CoadOrbit}
  \mathcal{O}_\rho = \{ \rho_g; g\in G\} \, .
\end{equation}
They also determine a map $J$ from the Lie algebra $\mathfrak{g}$
(cf.\ section~\ref{sect:coadorbits}) to functions on
$\mathcal{O}_\rho$ in which an element $A\in \mathfrak{g}$ is mapped
to a function $\mathcal{A} = J(A)$ with values
\begin{equation}
  \label{eq:ClassicalRepresentationJ}
  \mathcal{A} (g) = \langle g|\hat A|g\rangle = \rho_g(A) \, .
\end{equation}
The map $J$ is a classical representation of $\mathfrak{g}$.  For, if
$[\hat A,\hat B] = \rmi\hbar \, \hat C$, then the corresponding
functions $\mathcal{A}= J(A)$, $\mathcal{B} = J(B)$, and
$\mathcal{C}=J(C)$ satisfy a Poisson bracket relationship defined by
\begin{equation}
  \label{eq:PB}
  \{ \mathcal{A}, \mathcal{B} \}(g) =- \frac{\rmi}{\hbar} \langle g|[\hat
  A,\hat B]|g\rangle = \mathcal{C}(g) \, . 
\end{equation}

The Poisson bracket can be expressed in terms of local coordinates for
$\mathcal{O}_\rho$ as follows.  First observe that $\mathcal{O}_\rho$
is diffeomorphic to the coset space $H_\rho \backslash G$, where
\begin{equation}
  \label{eq:Hrho}
  H_\rho = \{ h\in G | \rho_h = \rho \} \, .
\end{equation}
Then the Lie algebra $\mathfrak{h}_\rho$ of the stability subgroup
$H_\rho$ is the set
\begin{equation}
  \label{eq:hLie}
  \mathfrak{h}_\rho = \{ X\in \mathfrak{g} \, |\, \rho([X,A])
  = 0 \, ,\forall\, A\in \mathfrak{g}\} \, .
\end{equation}
Thus, if $\{ A_i\}$ is a basis for $\mathfrak{h}_\rho$ and $\{
A_\nu\}$ completes a basis for $\mathfrak{g}$, coordinates are defined
for elements of $G$ about a point $g \in G$ by setting
\begin{equation}
  \label{eq:coordinatechart}
  g(\xi,x) = \rme^{-\frac{\rmi}{\hbar} \sum_i \xi^i A_i}
  \rme^{-\frac{\rmi}{\hbar} \sum_\nu x^\nu A_\nu} g \,. 
\end{equation}
It follows that at $g = g(0,0)$,
\begin{equation}
  \frac{\partial\mathcal{A}(g)}{\partial \xi^i} \equiv
  \frac{\partial\mathcal{A}(g(\xi,x))}{\partial \xi^i} \Big|_{\xi=x=0}
  =  -\frac{\rmi}{\hbar}\langle 0| [\hat A_i, \hat A(g)]|0\rangle = 0
  \, .
\end{equation}
However, the corresponding derivatives with respect to the $\{x^\nu\}$
coordinates do not, in general, vanish. Thus, the set $\{x^\nu\}$
serve as local coordinates for $\mathcal{O}_\rho \sim H_\rho
\backslash G$.  With respect to these coordinates, we then define
\begin{equation} 
  \label{eq:25} 
  (\partial_\mu \mathcal{A})(g) \equiv
  \frac{\partial\mathcal{A}(g(\xi,x))} {\partial x^\mu}
  \Big|_{\xi=x=0} =\sum_\nu\omega_{\mu\nu}A^\nu(g) \, , 
\end{equation}
where $A^\nu(g)$ is a coefficient in the expansion
\begin{equation}
  \label{eq:ExpansionOfX} 
  A(g) = \sum_i A^i(g) A_i + \sum_\nu A^\nu(g) A_\nu \, ,
\end{equation}
and
\begin{equation}
  \label{eq:omega}
  \omega_{\mu\nu} = -\frac{\rmi}{\hbar} \langle 0|[\hat A_\mu,\hat
  A_\nu]|0\rangle \, .
\end{equation}
The Poisson bracket of equation~(\ref{eq:PB}) is then expressed in the
familiar form
\begin{equation} \label{eq:PBracket}
  \{ \mathcal{A}, \mathcal{B} \}(g) = \sum_{\mu\nu} A^\mu(g)
  \omega_{\mu\nu} B^\nu(g)
  = \sum_{\mu\nu} (\partial_\mu \mathcal{A})(g)
  \omega^{\mu\nu} (\partial_\nu \mathcal{B})(g) \, ,
\end{equation}
where the matrix $(\omega^{\mu\nu})$ is defined such that
\begin{equation}
  \sum_\nu \omega^{\mu\nu} \omega_{\lambda\nu} = \delta^\mu_\lambda \, .
\end{equation}
In this form, the Poisson bracket can be extended to all $C^\infty$ 
functions on $\mathcal{O}_\rho$. 

It will be noted that the above results are expressed in terms of a
specific choice of the basis elements $\{ A_\nu\}$ of the Lie algebra.
Thus, it is important to ask if the results depend on the choice.  Two
distinct kinds of transformation of the basis are possible:
\begin{equation}
  A_\nu \to \sum_\mu A_\mu \gamma_{\mu\nu} \,.
\end{equation}
and
\begin{equation}
  A_\nu \to A_\nu + \sum_i A_i \gamma_{i\nu} \,.
\end{equation}
Combinations of the two kinds are also possible but it is instructive
to consider them separately. Transformations of the first kind leave
the subspace of the Lie algebra spanned by the elements $\{ A_\nu\}$
invariant. They generate normal coordinate transformations on
$\mathcal{O}_\rho$.  However, all physical expressions, like the
values of observables and their Poisson brackets, are manifestly
covariant relative to such coordinate transformation.  Transformations
of the second kind, corresponding to changes of the linear span of the
$\{A_\nu\}$ basis, are known as \emph{gauge transformations}.  Because
of the definition of the intrinsic subalgebra, the symplectic form
$\omega = \sum_{\mu\nu}\omega_{\mu\nu}\, \rmd x^\mu\, \rmd x^\nu$ is
seen to be invariant under a gauge transformation.  Moreover, from the
definition (\ref{eq:25}), it follows that
\begin{equation}
  \rmi\hbar(\partial_\nu \mathcal{A})(g) = \langle 0| [\hat A_\nu, \hat
  A(g)]|0\rangle \,.
\end{equation}
Thus, it follows that $(\partial_\nu \mathcal{A})(g)$ and hence the
Poisson bracket are also gauge invariant.

Note that different classical representations result from different
choices of the state $|0\rangle$,  even in the case when 
the representation $T$ is irreducible.

\subsection{Induced representations}   
\label{sect:ScalarReps}

Given an abstract unitary representation $T$ of the dynamical group
$G$ over a Hilbert space $\mathbb{H}$, a coherent state representation
$\Gamma$ is defined by specification of a functional $\langle\varphi
|$ on a $G$-invariant dense subspace $\mathbb{H}_D\subset\mathbb{H}$.
A state $|\psi\rangle \in\mathbb{H}_D $ then has coherent state wave
function $\psi$ defined over $G$ by
\begin{equation}
  \psi(g) = \langle \varphi |T(g) |\psi\rangle \, ,  
  \quad \forall\ g\in G \, .
\end{equation}
A Hilbert space $\mathcal{H}$ for a coherent state representation
$\Gamma$ is the completion of the space of such coherent state wave
functions with respect to the inner product
\begin{equation}
  (\psi,\psi') = \langle \psi |\psi'\rangle \, ,
\end{equation}
where the inner product on the right is that of $\mathbb{H}$.  The
coherent state representation $\Gamma$ is then defined by
\begin{equation}
   [\Gamma(g) \psi](g') = \psi (g'g) \,. 
\end{equation}
Clearly there are many coherent state representations depending on the
choice of the functional $\langle\varphi |$ and Hilbert space
$\mathbb{H}$.  If the functional $\langle\varphi |$ is chosen such
that
\begin{equation}
  \label{eq:Hconstraint}
  \psi(hg) = \langle \varphi |T(h)T(g) |\psi\rangle 
  = \chi(h) \psi(g) \, , \quad \forall\ h\in H_\rho \,, 
\end{equation}
where $\chi$ is a one-dimensional 
representation of the subgroup $H_\rho\subset G$, we say that the
coherent state representation $\Gamma$ is induced from the
representation $\chi$ of $H_\rho$.

If $G$ is compact, the space $\mathcal{H}$ of coherent state wave
functions is contained in $\mathcal{L}^2(G)$ and $\Gamma$ is
isomorphic to a subrepresentation of the regular
representation~\cite{Onofri} (this can be seen from the inner product
given in Section \ref{sect:InnerProducts}).  More generally, if
$\langle\varphi|$ is such that $\mathcal{H}$ consists of all functions
which satisfy equation~(\ref{eq:Hconstraint}) and whose absolute
values are in $\mathcal{L}^2(H_\rho\backslash G)$, then $\Gamma$ is
said to be a \emph{standard} (Mackey) induced
representation~\cite{mackey}.  In general, the representation $\Gamma$
is highly reducible.  It will be shown in section~\ref{subsec:Irreps}
that, by imposing extra conditions on the choice of the functional
$\langle\varphi |$, it is possible to proceed directly to the
irreducible representations of quantization.

The above coherent state wavefunctions are defined as functions over
the group $G$.  However, in practical applications, it is generally
more useful to represent them as functions over a suitable set of
$H_\rho\backslash G$ coset representatives~\cite{Onofri}.  Recall that
a set of coset representatives $K=\{ k(g)\in H_\rho g ; g\in G\}$
defines a unique factorization $g = h(g) k(g)$, with $h(g)\in H_\rho$,
of every $g\in G$.  Hence, it follows from the identity
(\ref{eq:Hconstraint}) that the restriction of $\psi \in \mathcal{H}$
to the subset $K\subset G$ is sufficient to uniquely define $\psi$.
Often it is also convenient to consider factorizations of the type $g
= h(g)k(g)$ with $h(g) \in H_{\rho}^c$ and $k(g) \in K$, where $K$ is
a subset of $H_{\rho}^c\backslash G^c$ coset representatives and
$H_{\rho}^c$ and $G^c$ are the complex extensions of $H_{\rho}$ and
$G$, respectively.  Note that the identity (\ref{eq:Hconstraint}) does
not require that the wavefunctions are functions on $H_\rho \backslash
G$; in general, $\psi$ need only be a section of a complex line bundle
associated to the principal bundle $G \to H_\rho \backslash G$.  The
Hilbert space $\mathcal{H}$, then, can be viewed as a space of
sections of a complex line bundle over $H_\rho \backslash
G$~\cite{HarmAnalysis}.

We now show that, if a representation $\Gamma$ is induced from a
representation $\chi$ of $H_\rho$ having the property that
\begin{equation}
   \rmi\frac{\rmd}{\rmd t} \chi(\rme^{-\rmi At})\Big|_{t=0} = 
   \chi(A) \equiv\rho(A)\, , \quad A\in \mathfrak{h}_\rho\, ,
\end{equation} 
then the corresponding representation of the Lie algebra 
$\mathfrak{g}$, defined by
\begin{equation}
  \label{eq:CSRepresentationOfAlgebra}
  [\Gamma(A)\psi](g) = \langle \varphi | T(g) \hat{A} |\psi\rangle
  = \psi(gA), \quad A \in \mathfrak{g} \, ,
\end{equation}
where 
\begin{equation}
  \psi(gA) = \rmi\frac{\rmd}{\rmd t}\psi(g\rme^{-\rmi tA})\Big|_{t=0} \,,
\end{equation}
is a subrepresentation of that given by prequantization.  (Such a
relationship was shown in more restricted contexts, for example, by
Dunne~\cite{dunne} and Rawnsley~\cite{Rawnsley}.)

Note, however, that for this prequantization 
to be possible, the representation $\chi$ of $\mathfrak{h}_\rho$
defined by $\rho$ must be a subrepresentation of the restriction to
$\mathfrak{h}_\rho \subset \mathfrak{g}$ of some 
unitary representation $T$ of $\mathfrak{g}$.  When this condition is
satisfied, we say (in the language of geometric quantization) that the
classical representation of $\mathfrak{g}$ defined by $\rho$ is {\em
  quantizable}.

First observe that 
\begin{equation}
  \psi(gA) =\psi(A(g)g), 
\end{equation}
where $A(g) = {\rm Ad}_g(A)$.  Substitution of the identities
\begin{equation}
  \label{eq:39}
  \psi (A_i g)= \chi (A_i) \psi (g) , \qquad \psi(A_\nu g) = \rmi\hbar
  (\partial_\nu \psi)(g)\,, 
\end{equation}
where
\begin{equation}
  \rmi\hbar (\partial_\nu \psi)(g) = \rmi\hbar\frac{\partial}{\partial
  x^\nu} \psi( \rme^{-\frac{\rmi}{\hbar} \sum_\nu x^\nu  A_\nu}
  g)\Big|_{x=0}\,,
\end{equation}
into the expansion equation~(\ref{eq:ExpansionOfX}) of $A(g)$, then gives
the explicit expression
\begin{equation}
  [\Gamma(A)\psi](g) = \sum_i A^i(g) \chi (A_i) \psi(g)
  + \rmi \hbar \sum_\nu A^\nu(g) (\partial_\nu 
  \psi)(g) \, 
  \label{eq:CSrep} 
\end{equation}
for the action of $\Gamma(A)$ on coherent state wave functions.

The expression (\ref{eq:CSrep}) of $\Gamma$ depends on the expansion
(\ref{eq:ExpansionOfX}) of $A(g)$.  Thus it is coordinate-dependent
and gauge-dependent.  However, it can be expressed in a covariant
form by taking advantage of the symplectic structure of the classical
phase space.  By expanding the classical representation
$\mathcal{A}(g) = \rho( A(g))$ of $A\in \mathfrak{g}$,
\begin{equation}
  \mathcal{A}(g) = \sum_i A^i(g) \chi (A_i) + \sum_\nu A^\nu(g)
  \rho(A_\nu) \,,
\end{equation}
equation~(\ref{eq:CSrep}) becomes 
\begin{equation} 
  \label{eq:3.30}
  [\Gamma(A)\psi](g) = \mathcal{A}(g)\psi(g)  +  \rmi\hbar \sum_\nu
  A^{\nu}(g) (\nabla_\nu \psi)(g) \, ,
\end{equation}
where
\begin{equation}
  \rmi\hbar (\nabla_\nu \psi)(g) = \rmi\hbar (\partial_\nu
  \psi)(g)- \rho(A_\nu)\psi(g) \,.
\end{equation}
The first term, $\mathcal{A}(g)\psi(g)$, of equation~(\ref{eq:3.30})
is manifestly covariant.  Moreover, from the definition,
$[\Gamma(A)\psi](g) = \psi(A(g)g)$, it follows that
\begin{equation}
  \rmi\hbar \sum_\nu A^{\nu}(g) (\nabla_\nu \psi)(g) 
  = \psi (A(g)g) - \rho(A(g))\psi(g)\,. 
\end{equation}
Thus, the second term is also covariant.  We shall refer to the
operator $\nabla_{X_{\mathcal{A}}}=\sum_\nu A^\nu\nabla_\nu$, defined
such that
\begin{equation}
  [\nabla_{X_{\mathcal{A}}}\psi](g) = \sum_\nu  A^{\nu}(g) (\nabla_\nu
  \psi)(g)\,,
\end{equation}
as a covariant derivative, in accordance with standard terminology.
The induced representation $\Gamma(A)$ of an arbitrary element $A\in
\mathfrak{g}$ is then expressed in the covariant form
\begin{equation}
  \label{eq:ScalarPrequantization}
  \Gamma(A) = \mathcal{A} + \rmi\hbar \nabla_{X_{\mathcal{A}}} \,.
\end{equation} 
A similar expression can also be derived within the framework of
Berezin quantization \cite{Rawnsley}.

It is shown in the appendix that, for a particular choice of gauge,
$\nabla_{X_{\mathcal{A}}}$ is expressible as a sum
\begin{equation}
  \nabla_{X_{\mathcal{A}}} =X_{\mathcal{A}} + \frac{\rmi}{\hbar}
  \theta(X_{\mathcal{A}}) \,,
\end{equation}
where $X_{\mathcal{A}}$ is a Hamiltonian vector field defined by the
classical function $\mathcal{A}$ and $\theta$ is a one-form (gauge
potential). It is also shown that the symplectic two-form $\omega$ of
the manifold $\mathcal{O}_\rho$ is the exterior derivative
\begin{equation}
  \omega = \rmd\theta \,.
\end{equation}
Thus, $\theta$ is a symplectic potential for $\mathcal{O}_\rho$.  Note
that the values of the symplectic potential $\theta$ depend on the
choice of $\{ A_\nu\}$ (cf.\ equation~(\ref{eq:coordinatechart})).  As
a consequence, $\theta$ is only defined to within a gauge
transformation $\theta \to \theta + \rmd\alpha$.  However, its
exterior derivative is gauge independent.

The coherent state representation $\Gamma : A \to \mathcal{A} +
\rmi\hbar\nabla_{X_{\mathcal{A}}}$ is now observed to be of the
standard form of prequantization in the theory of geometric
quantization.  Thus, prequantization of an algebraic model is
equivalent to a standard (Mackey) representation induced from a
one-dimensional irrep of a suitable subgroup.  However, depending on
the choice of functional $\langle \varphi|$ and starting Hilbert space
$\mathbb{H}$, the Hilbert space $\mathcal{H}$ of a coherent state
representation may be an invariant subspace of that of
prequantization.  Indeed, as we show in the following section, it is
often possible to choose the functional $\langle \varphi|$ such that
the coherent state representation is irreducible.

Geometric quantization shows that prequantization can be extended to
the whole infinite dimensional classical algebra of all functions on
the phase space.  However, the extension is not fully reducible and,
as presently formulated, does not apply generally to an arbitrary
coherent state representation, i.e., a coherent state representation
that is not equivalent to a standard induced representation.  The
theory of induced representations also extends, albeit in different
ways.  For example, it is possible to induce coherent state
representations from a representation of a subgroup $H\subset G$ for
which $H\backslash G$ is not symplectic. It is also possible to induce
representations that are not unitary.  Some of the possibilities will
be illustrated with examples in section~\ref{sec:ScalarCSExamples}.

\subsection{Irreducible representations}
\label{subsec:Irreps}

The full quantization of an algebraic model corresponds to
construction of an irreducible unitary representation of its SGA.  The
usefulness of scalar coherent state induction, and indeed the full VCS
theory, resides in its facility to construct such representations in a
practical and computationally tractable manner.  
All that is needed is a functional $\langle \varphi|$ 
that uniquely characterizes an irrep.  Such a functional 
can often be defined, for example, by extending the
condition (\ref{eq:Hconstraint}) to a suitable subgroup $P$ in the
chain $H_\rho\subset P\subset G^c$, where $G^c$ is the complex
extension of $G$.  In making the extension from a real
subgroup $H_\rho\subset G$ to a complex subgroup $P\subset G^c$, we recall
that a similar extension \cite{streater,dunne}, applied to Kirillov's
orbit method, resulted in a  method for constructing certain irreducible
representations for semisimple Lie groups.

In making this extension, a technical concern is that, whereas a
representation of a SGA $\mathfrak{g}$ extends linearly to the complex
extension $\mathfrak{g}^c$ of $\mathfrak{g}$, the corresponding
extension of the generic unitary representation $T$ of the real group
$G$ may not converge for all of $G^c$.  However, it is sufficient for
the purpose of defining an irreducible coherent state representation
if the action of $T$ on some dense subspace $\mathbb{H}_D \subseteq
\mathbb{H}$ can be extended to a suitable subset $U(P)\subset P$ of a
subgroup $P\subset G^c$ which contains $H_\rho$. Let $\tilde\chi$
denote a one-dimensional irrep of $P\subset G^c$ which restricts to a
unitary irrep $\chi$ of $H_\rho\subset P$. Now suppose a functional
$\langle \varphi |$ is chosen such that
\begin{equation}
  \label{eq:Nconstraint}
  \langle \varphi |T(z)T(g) |\psi\rangle 
  = \tilde\chi(z) \psi(g) \, , \quad \forall\ z\in U(P) \,. 
\end{equation}
It will be shown by examples in the following sections that, for many
categories of groups, there are natural choices of $P$ and its
representation $\tilde\chi$ for which the corresponding coherent state
representation is irreducible.

Subgroups satisfying these conditions are familiar in the holomorphic
induction of irreducible representations~\cite{Harish-Chandra}.  For
example, if $G$ were semisimple and the isotropy subgroup $H_\rho
\subset G$ for the coadjoint orbit $H_\rho\backslash G$, as defined
above, were a Cartan subgroup, then a suitable subgroup $P\subset G^c$
would be the Borel subgroup generated by $H_\rho$ and the exponentials
of a set of raising (or lowering) operators.  A suitable
one-dimensional representation of $P$ would then be defined by a
dominant integral highest weight for a unitary irrep of $G$.  More
generally, if $H_\rho$ were a Levi subgroup, $P$ would be parabolic.
Non-unitary irreps can also be induced in this way.  However, they are
not normally described as quantizations.

The construction outlined above is a generalization of holomorphic
induction and, for convenience, in this situation we will speak of
``the representation of $G$ induced from a representation of a
subgroup $P \subset G^c$.'' 

Apart from imposing the stronger condition (\ref{eq:Nconstraint}), the
coherent state construction is the same as in
section~\ref{sect:ScalarReps}.  However, the stronger condition
restricts the set of coherent state wave functions to a subset with
the result that the coherent state representation becomes an
irreducible subresentation of that given by prequantization.

Now if a unitary coherent state representation $\Gamma$ of a dynamical
group $G$ induced from a representation $\tilde \chi$ of a subgroup
$P\subset G^c$ defines an irreducible representation of the Lie
algebra $\mathfrak{g}$ and if the representation $\tilde\chi$
satisfies the equality
\begin{equation}
  \rmi\frac{\rmd}{\rmd t} \tilde\chi(\rme^{-\rmi At})\Big|_{t=0} =
  \tilde\chi(A)  \equiv 
  \rho(A)\, ,  \quad A\in \mathfrak{p}\, ,
  \label{eq:QCondition}
\end{equation} 
then we say that $\Gamma$ is a quantization of the classical
representation of $\mathfrak{g}$ defined by $\rho$.

Note, however, that for this quantization to be possible the
representation $\chi$ of $\mathfrak{h}$ must extend to a
representation $\tilde\chi$ of a subalgebra $\mathfrak{p}
\subset\mathfrak{g}^c$ which is contained in a unique irrep of
$\mathfrak{g}^c$ which restricts to a unitary irrep of $\mathfrak{g}$.

In the theory of geometric quantization, one would say that the choice
of subgroup $P\subset G^c$ defines an \emph{invariant
  polarization}~\cite{kostant,auslander}.  Recall that a basis for
$\mathfrak{h}^c\backslash\mathfrak{g}^c$ defines a basis for the
complex extension of the tangent space at every point of the classical
phase space $H_\rho\backslash G$.  A polarization provides a
separation of the tangent space at each point of this phase space into
canonical space-like and momentum-like subspaces.

Let $\mathfrak{p}$ denote the Lie algebra of $P$.  According to
Woodhouse~\cite{woodhouse}, the subalgebra $\mathfrak{p}\subset
\mathfrak{g}^c$ generates an invariant polarization if it satisfies
the conditions:
\begin{enumerate}
\item[(i)] $\rho([A,B])=0$ for any $A,B \in \mathfrak{p},$
\item[(ii)] ${\rm dim}_{\mathbb{R}}\, \mathfrak{g} +
  {\rm dim}_{\mathbb{R}}\, \mathfrak{h}_\rho = 2 
  \,{\rm dim}_{\mathbb{C}}\, \mathfrak{p},$
\item[(iii)] $\mathfrak{p}$ is invariant under the adjoint action 
  of $H_\rho$.
\end{enumerate}
The first condition ensures that the polarization is isotropic, i.e.,
contains no canonically conjugate observables.  The second condition
ensures that $\mathfrak{p}$ is a maximal subalgebra for which the
first condition holds; the polarization is then said to be Lagrangian
on $H_\rho \backslash G.$ This condition ensures that $\mathfrak{p}$
is sufficiently large that a representation of the group $P$
characterizes an irrep of $G.$ The final condition ensures that the
polarization is well-defined on $H_\rho \backslash G.$ These
conditions extend the definition of a parabolic subalgebra for a
semisimple Lie algebra to the general situation.

As we illustrate with several examples in
section~\ref{sec:ScalarCSExamples}, the choice of a suitable subgroup
$P\subset G^c$ for a coherent state quantization also defines an
invariant polarization according to the above criteria.

\subsection{Coherent state inner products} \label{sect:InnerProducts}

For a coherent state irrep that belongs to the discrete series, an
inner product is defined in the following standard way.  For a given
reference state $|\varphi\rangle$, let $\mathbb{I}$ denote the
integral
\begin{equation}
  \label{eq:ResolutionOfIdentity}
  \mathbb{I} = \int T(g^{-1}) |\varphi\rangle\langle\varphi|T(g)  \,
  \rmd v(g) \, , 
\end{equation}
where $\rmd v$ is a right-invariant measure on $G$.  This integral
converges if $|\varphi\rangle$ is a normalizable state vector in an
irreducible subspace of $T$ that carries a discrete series
representation.  The integral then defines $\mathbb{I}$ as a
well-defined operator on the Hilbert space.  Moreover, it commutes
with the representation $T(g')$ of any element $g'\in G$.  Hence, by
Schur's lemma, $\mathbb{I}$ is a multiple of the identity on the
irreducible subspace containing the vector $|\varphi\rangle$.  It
follows that the space of coherent state wave functions
\begin{equation}
  \mathcal{H} = \Bigl\{ \psi \Bigl| \psi(g) = \langle \varphi
  |T(g)|\psi\rangle ,\; |\psi\rangle \in \mathbb{H} \Bigr\}\, ,
\end{equation}
where $\mathbb{H}$ is the Hilbert space for the representation $T$,
has inner product given to within a convenient norm factor by
\begin{equation}
  \label{eq:NormOfVector}
  (\psi,\psi') = \int \psi^* (g) \psi'(g)\, \rmd v(g) \, .
\end{equation}

If the representation $\chi$ of $H_\rho$ is unitary, the coherent
state wave functions have the property
\begin{equation}
  \psi^* (hg) \psi'(hg) =\psi^* (g) \psi'(g) \,.
\end{equation}
Then the integral over the group in equation~(\ref{eq:NormOfVector})
can be restricted to an integral over the coset space $H_\rho
\backslash G$ with a right invariant measure induced from that on $G$.

When $\langle\varphi|$ is a functional on a dense subspace
$\mathbb{H}_D$ of $\mathbb{H}$, the integral $\mathbb{I}$ may not
converge.  However, the corresponding integral over $H_\rho \backslash
G$ may converge and, if so, it is sufficient to define an inner
product of coherent state wave functions in parallel with Mackey's
construction of inner products for induced representations of
semi-direct product groups.

Inner products for more general coherent state representations are
constructed by K-matrix methods~\cite{rowe88} and the related
integral methods of Rowe and Repka~\cite{triplets}.

\section{Examples}\label{sec:ScalarCSExamples}

Examples are given in the following to illustrate systematic
procedures for carrying out the prescriptions of geometric
quantization within the framework of coherent state representation
theory.  The first example, for the nilpotent Heisenberg-Weyl algebra,
serves as a useful prototype for more general applications.  The
familiar quantizations of this algebra known as Schr\"odinger
quantization and the Bargmann-Segal representation~\cite{BS} are both
illustrated.  The second and third examples are prototypes of
semisimple and semidirect sum Lie algebras, respectively.  The
examples show that coherent state theory provides simple and natural
routes through the (sometimes subtle) methods of geometric
quantization.  Often there is more than one path. There may be a
choice of polarization (as illustrated by two representations, one
real and one holomorphic, for the HW algebra) and a choice of the
functional form of the resulting Hilbert space (illustrated for
$SU(2)$).

\subsection{The nilpotent Heisenberg-Weyl (HW) algebra}
\label{subsubsec:Schrodinger}

A generic unitary representation $T$ of the HW algebra is spanned by
Hermitian operators, $\{ \hat{q}, \hat{p}, \hat{I} \},$ on a Hilbert
$\mathbb{H}$ with commutation relations
\begin{equation}
  \label{eq:hw(3)CommutationRelations}
  [\hat{q}, \hat{p}] = \rmi \hbar\, \hat{I} \, , \qquad
  [\hat{q},\hat{I}] = 0 \, , \qquad [\hat{p}, \hat{I}] = 0 \, .
\end{equation}

The representations of the so-called oscillator group, which contains
the Heisenberg-Weyl group as a normal subgroup, were constructed by
Streater~\cite{streater} using both Mackey and Kirillov methods.

\subsubsection{Schr\"odinger quantization} 

Let $|0\rangle \in \mathbb{H}$ denote any normalized state for 
which
\begin{equation}
\label{eq:hw(3)NormalizedState}
   \langle 0| \hat I |0\rangle = 1 \, , \qquad
   \langle 0| \hat q |0\rangle = 
   \langle 0| \hat p |0\rangle = 0 \, .
\end{equation}
If elements of the HW group are parameterized
\begin{equation}
  T(g(\theta,q,p)) = \rme^{-\frac{\rmi}{\hbar} \theta \hat{I}}
  \rme^{-\frac{\rmi}{\hbar} p \hat{q} } \rme^{\frac{\rmi}{\hbar} q
  \hat{p}} \, ,
\end{equation}
the group conjugates of $\{ \hat{q}, \hat{p}, \hat{I} \}$ are
\begin{equation} 
  \eqalign{
  \hat q(g) = T(g)\, \hat q \, T(g^{-1}) = \hat q + q\hat I \, , \\
  \hat p(g) = T(g)\, \hat p \, T(g^{-1}) = \hat p + p\hat I \, , \\
  \hat I(g) = T(g)\, \hat I \, T(g^{-1}) = \hat I \, . }
\end{equation}
and a classical realization of the HW algebra is given by the
functions $\{ \mathcal{Q}, \mathcal{P}, \mathcal{I} \}$ of $p$ and $q$
with
\begin{equation}
  \eqalign{
  \mathcal{Q}(p,q) = \langle 0| \hat q(g) |0\rangle = q \, , \\
  \mathcal{P}(p,q) = \langle 0| \hat p(g) |0\rangle = p \, , \\
  \mathcal{I}(p,q) = \langle 0| \hat I(g) |0\rangle = 1 \, . }
\end{equation}
The Poisson bracket of $\mathcal{Q}$ and $\mathcal{P}$, for example,
is given by
\begin{equation} 
   \{ \mathcal{Q},\mathcal{P}\}(p,q) = -\frac{\rmi}{\hbar}\langle 0|
   [\hat q(g),\hat p(g)]|0\rangle = \mathcal{I}(p,q) \, .
\end{equation}

An induced representation of the HW algebra, equivalent to a
prequantization, is now constructed by coherent state techniques as
follows.  (Note that we employ a different coordinate chart than used
in section~\ref{sect:ScalarReps}; although equivalent results are
obtained in any chart, the coordinates used here are standard for this
example.)  Choosing $\langle\varphi|=\langle 0|$ to be some normalized
state satisfying equation~(\ref{eq:hw(3)NormalizedState}), as above,
and factoring out the phases generated by the identity $\hat I$, a
state $|\psi\rangle$ of a model with the HW algebra as its SGA is
assigned a coherent state wave function $\psi$ defined over the
classical phase space (the $p-q$ plane) by
\begin{equation}
   \label{eq:pqCoords}
  \psi(p,q) = \langle 0|  \rme^{-\frac{\rmi}{\hbar} p \hat{q} }
  \rme^{\frac{\rmi}{\hbar} q \hat{p}} |\psi\rangle \, .
\end{equation}
The corresponding coherent state representation $\Gamma$ of an element
$\hat{A}$ of the Heisenberg-Weyl algebra, defined generally by
\begin{equation}
  \eqalign{
  [\Gamma(\hat{A})\psi](p,q) &= \langle 0|\rme^{-\frac{\rmi}{\hbar} p
  \hat{q}} \rme^{\frac{\rmi}{\hbar} q \hat{p}}\hat{A} | \psi \rangle \\
  &= \langle 0|  \rme^{-\frac{\rmi}{\hbar} p \hat{q} }
  \bigl( \hat{A} + \frac{\rmi}{\hbar} q[\hat{p},\hat{A}] \bigr)
  \rme^{\frac{\rmi}{\hbar} q \hat{p}} |\psi\rangle \, ,}
\end{equation}
is then the induced representation
\begin{equation}
  \Gamma(\hat{q}) = q + \rmi\hbar\frac{\partial}{\partial p}\,  , \qquad
  \Gamma(\hat{p}) = -\rmi \hbar \frac{\partial}{\partial q}\, ,\qquad
  \Gamma(\hat{I}) = 1 \, . 
  \label{eq:inducedHW}
\end{equation}

This representation is obtained in geometric quantization starting
with the Poisson bracket of any two functions $\mathcal{A}$ and
$\mathcal{B}$ in the HW algebra in the form
\begin{equation} 
  \{ \mathcal{A},\mathcal{B}\} =
  \frac{\partial\mathcal{A}}{\partial q}
  \frac{\partial\mathcal{B}}{\partial p}  
  - \frac{\partial\mathcal{A}}{\partial p}
  \frac{\partial\mathcal{B}}{\partial q} \, .
\end{equation}
Corresponding vector fields are then defined by
\begin{equation} 
  X_{\mathcal{Q}} = \frac{\partial}{\partial p}\, ,\qquad 
  X_{\mathcal{P}} = -\frac{\partial}{\partial q}\, , \qquad 
  X_{\mathcal{I}} = 0 \, ,
\end{equation}
and a symplectic form, for which
\begin{equation}
  \{ \mathcal{A},\mathcal{B}\}=\omega(X_\mathcal{A},X_\mathcal{B}) \, ,
\end{equation}
is given by
\begin{equation}
  \omega = \rmd q\wedge \rmd p \, .
\end{equation}
This two-form is exact and expressible as the exterior derivative
$\omega = \rmd \theta$ of a variety of one-forms.  For the $(p,q)$
coordinates defined by setting $T(g(p,q)) = \rme^{-\frac{\rmi}{\hbar}
  p \hat{q} } \rme^{\frac{\rmi}{\hbar} q \hat{p}}$, cf.\ 
equation~(\ref{eq:pqCoords}), the identities
\begin{equation}
  \eqalign{
  \hat q T(g(p,q)) 
  = \rmi\hbar \frac{\partial}{\partial p} T(g(p,q)) \,, \\
  \hat p T(g(p,q)) 
  =(-\rmi\hbar \frac{\partial}{\partial q} +p) T(g(p,q)) \,, }
\end{equation}
imply that an appropriate one-form is
\begin{equation}
  \theta = -p\, \rmd q \, .
\end{equation}
Prequantization of the $\{ \mathcal{Q},\mathcal{P},\mathcal{I}\}$
functions then gives
\begin{equation}
  \eqalign{
  \Gamma(\mathcal{Q}) = \mathcal{Q} + \rmi\hbar X_{\mathcal{Q}} -
  \theta(X_{\mathcal{Q}}) &= q +
  \rmi\hbar \frac{\partial}{\partial p} \, , \\
  \Gamma(\mathcal{P}) = \mathcal{P} + \rmi\hbar X_{\mathcal{P}} -
  \theta(X_{\mathcal{P}}) &= -
  \rmi\hbar \frac{\partial}{\partial q} \, ,  \\
  \Gamma(\mathcal{I}) = \mathcal{I} -X_{\mathcal{I}} -
  \theta(X_{\mathcal{I}}) &= 1 \, , }
\end{equation}
which is identical to the induced representation of
equation~(\ref{eq:inducedHW}).  

To obtain an irreducible representation, a functional $\langle\varphi|$
on a suitably-defined dense subspace  $\mathbb{H}_D \subset \mathbb{H}$
may be chosen such that
\begin{equation}
  \label{eq:PropertiesOfDeltaFunctionState}
  \langle\varphi|\hat q |\psi\rangle = 0\, , \quad \forall \
  |\psi\rangle \in \mathbb{H}_D \, .
\end{equation}

This choice corresponds to choosing the real polarization
$\mathfrak{p}$ spanned by the operators $\hat q$ and $\hat I$.  Note
that the state $|\varphi\rangle$ is not a normalizable state vector of
the Hilbert space of square-integrable functions on the HW group.
Nevertheless, the bra vector $\langle\varphi|$ is a well-defined
functional on $\mathbb{H}_D$.  The coherent state wave functions, for
states in this dense subspace, are then the $p$-independent
functions, given by
\begin{equation}
  \label{eq:hw(3)CSWavefunction}
  \psi(q) = \langle\varphi| \rme^{\frac{\rmi}{\hbar} q \hat{p}} | \psi
  \rangle \, ,
\end{equation}
and the coherent state representation of the algebra reduces to the
familiar irreducible Schr\"odinger representation
\begin{equation}
  \Gamma(\hat{q}) = q \, , \qquad
  \Gamma(\hat{p}) = -\rmi \hbar \frac{\partial}{\partial q} \, ,\qquad
  \Gamma(\hat{I}) = 1 \, .
\end{equation}

\subsubsection{The Bargmann-Segal representation}

To obtain a classical Bargmann-Segal representation of the HW
algebra, choose any normalized state $|0\rangle$ in the Hilbert space
such that
\begin{equation}
  \langle 0| \hat{a}^\dagger|0\rangle = \langle 0|\hat{a}|0\rangle =0
  \, , \qquad \langle 0|\hat I|0\rangle =1 \, ,
\end{equation}
where
\begin{equation}
   \hat{a}^\dagger = \frac{1}{\sqrt{2}}
  \Big(\hat{q}-\frac{\rmi}{\hbar}\hat{p}\Big) \, ,\qquad 
   \hat{a} = \frac{1}{\sqrt{2}}
  \Big(\hat{q}+\frac{\rmi}{\hbar}\hat{p}\Big) \, .
\end{equation}
With group elements parameterized by the
factorization
\begin{equation}
   T(g(z,z^*,\varphi)) = \rme^{(\rmi\varphi - \frac{1}{2} |z|^2)\hat{I}}
   \rme^{-z^* \hat{a}^\dagger} \rme^{z\hat{a}} \, ,
\end{equation}
the group conjugates of $\{ \hat{a}^\dagger, \hat{a}, \hat{I} \}$ are
given by
\begin{equation} 
  \eqalign{
  \hat{a}^\dagger(z,z^*) =\: T(g)\, \hat{a}^\dagger \, T(g^{-1}) 
   &=\: \hat{a}^\dagger + z\hat I \, , \\
  \hat{a}(z,z^*) = T(g)\, \hat{a} \, T(g^{-1}) &=\: \hat{a} + z^*\hat I
  \,, \\
  \hat I(z,z^*) = T(g) \, \hat I \, T(g^{-1}) &=\: \hat I \, , }
\end{equation}
and the corrresponding conjugates of $\hat q$ and $\hat p$ are
\begin{equation}
  \eqalign{
  \hat{q}(z,z^*) = \hat{q} + \frac{1}{\sqrt{2}} (z+z^*)\hat I   
  \,, \\
  \hat{p}(z,z^*) = \hat{p} +  \frac{\rmi\hbar}{\sqrt{2}} (z-z^*) \hat I
  \,.}
\end{equation}
This parametrization leads to a classical
realization of the HW algebra in which $\{ \hat{q},\hat{p},\hat{I}\}$
map to functions $\{ \mathcal{Q}, \mathcal{P}, \mathcal{I}\}$ of $z$
and $z^*$ defined by
\begin{equation}
  \eqalign{
  \mathcal{Q}(z,z^*) = \langle 0| \hat{q}(z,z^*) |0\rangle &=
  \frac{1}{\sqrt{2}} (z+z^*) \, , \\
  \mathcal{P}(z,z^*) = \langle 0| \hat{p}(z,z^*) |0\rangle &=
  \frac{\rmi\hbar}{\sqrt{2}} (z-z^*) \, ,  \\
  \mathcal{I}(z,z) = \langle 0| \hat I(z,z^*) |0\rangle &= 1 \,,}
\end{equation}
and with Poisson bracket given, for example, by
\begin{equation} 
  \{ \mathcal{Q},\mathcal{P}\}(z,z^*) = -\frac{\rmi}{\hbar}\langle
  0|[\hat{q}(z,z^*),\hat{p}(z,z^*)]|0\rangle = \mathcal{I}(z,z^*) \, .
\end{equation} 

With $\langle\varphi|=\langle 0|$, a state $|\psi\rangle$ is now
assigned a coherent state wave function $\psi$ defined over the
complex $z$ plane by
\begin{equation}
  \psi(z,z^*) = \langle\varphi|  \rme^{-z^*\hat{a}^\dagger}
  \rme^{z\hat{a}} |\psi\rangle \, .
\end{equation}
The corresponding coherent state representation $\Gamma$ of an element
$\hat{A}$ of the Heisenberg-Weyl algebra, defined generally by
\begin{equation}
  [\Gamma(\hat{A})\psi](x) = \langle\varphi| \rme^{-z^*\hat{a}^\dagger}
  \rme^{z\hat{a}}\hat{A} | \psi \rangle
  = \langle\varphi|  \rme^{-z^*\hat{a}^\dagger} (\hat{A} +
  z[\hat{a},\hat{A}]) \rme^{z\hat{a}} |\psi\rangle \, ,
\end{equation}
then gives the prequantization
\begin{equation}
  \Gamma(a) = \frac{\partial}{\partial z} \, , \qquad
  \Gamma(a^\dagger) =z  - \frac{\partial}{\partial z^*} \, ,\qquad
  \Gamma(I) = 1 \, . 
\end{equation} 

To obtain an irreducible representation, define $|\varphi\rangle$ to
be the vacuum state for which
\begin{equation}
   \hat{a} |\varphi\rangle = 0 \, , \qquad \hat{I}|\varphi\rangle =
   |\varphi\rangle \, .
\end{equation}
This state satisfies the equation
\begin{equation}
  \langle\varphi|\hat I|\psi\rangle = \langle\varphi|\psi\rangle \, ,\qquad
  \langle\varphi|\hat{a}^\dagger|\psi\rangle = 0 \, ,
\end{equation}
and defines a complex polarization $\mathfrak{p}\subset
\mathfrak{g}^c$ spanned by the operators $\hat{a}^\dagger$ and $\hat
I$.  The coherent state wave functions are now the holomophic
functions, given for $|\psi\rangle$ in the dense subspace of
$\mathbb{H}$ generated by the action of finite powers of $a^\dagger$
on the vacuum state by
\begin{equation}
  \psi(z) = \langle\varphi|  \rme^{z\hat{a}} | \psi \rangle \,.
\end{equation}
The corresponding coherent state representation of the HW algebra is
now the well-known Bargmann-Segal representation
\begin{equation}
  \Gamma(a) = \frac{\partial}{\partial z} \, , \qquad
  \Gamma(a^\dagger) =z \, ,\qquad
  \Gamma(I) = 1 \, , 
\end{equation}
which is known to be irreducible.

The Hilbert $\mathcal{H}$ space for this irrep is inferred in coherent
state theory from the requirement that $\partial /\partial z$ should
be the Hermitian adjoint of $z$ for a unitary representation.  Thus,
$\mathcal{H}$ has an orthonormal basis $\{ \psi_n ; n=0,1,2, \ldots
\}$ with
\begin{equation}
  \psi_n (z) = \frac{z^n}{\sqrt{n!}} 
\end{equation}
and inner product
\begin{equation}
  (\psi_m,\psi_n) =  \frac{1}{\sqrt{m!n!}}
  \left.\left(\frac{\partial^m}{\partial
  z^m} z^n \right)\right|_{z=0} = \delta_{mn} \,.
\end{equation} 
From the observation that
\begin{equation}
  \int \left( \frac{\partial\psi}{\partial z}\right)^* \rme^{-zz^*}
  \psi'(z) \, \rmd z\, \rmd z^* =
  \int \psi(z)^* \rme^{-zz^*} z \psi'(z) \, \rmd z\, \rmd z^* \,,
\end{equation}
it is also determined that $\mathcal{H}$ is the space of holomorphic
functions with norm
\begin{equation}
  (\psi,\psi) = \frac{1}{2\pi} \int |\psi(z)|^2 \rme^{-zz^*} \, \rmd
  z\, \rmd z^* \,.
\end{equation}
This Hilbert space $\mathcal{H}$ is the well-known Bargmann-Segal
space of entire analytic functions.
 
\subsection{The semisimple $su(2)$ algebra}

Representations of the groups $SO(3)$ and $SO(2,1)$ were constructed
within the framework of Kirillov's orbit method by Dunne~\cite{dunne}.
We consider here only $su(2)$ (isomorphic to $so(3)$) as an example of
a semisimple Lie algebra.

Suppose the regular representation $T$ of the $su(2)$ algebra is
spanned by three components of angular momentum
$(\hat{S_1},\hat{S_2},\hat{S_3})$ with commutation relations
\begin{equation} 
  [\hat{S_i},\hat{S_j}] = \rmi \hat{S_k}\, ,\quad i,j,k\; {\rm cyclic,}
\end{equation}
acting on $\mathcal{L}^2(SU(2))$.

Elements of the SU(2) group can be parameterized in many ways.  The
standard parameterization, in terms of Euler angles,
\begin{equation}
  T(g(\alpha,\beta,\gamma)) = \rme^{-\rmi \alpha \hat{S_3}}
  \rme^{-\rmi \beta \hat{S_2}} \rme^{-\rmi\gamma \hat{S_3}} \, ,
\end{equation} 
leads to a classical realization of the $su(2)$ Lie algebra and a
prequantization.  However, with this parameterization, it is not so
easy to identify a polarization and an irreducible subrepresentation.
Parameterizations that lead naturally to irreducible quantizations are
defined as follows.

\subsubsection{Representation by functions on a circle}

Because the Lie algebras $su(2)$ and $sl(2,\mathbb{R})$ are both real
forms of $sl(2,\mathbb{C})$, it follows that the irreps of $su(2)$ are
defined by corresponding finite-dimensional irreps of
$sl(2,\mathbb{R})$.  Thus, it is useful to regard the operators $\{
\hat{S}_1, \rmi\hat{S}_2, \hat{S}_3\}$ as spanning a
finite-dimensional irrep of $sl(2,\mathbb{R})$ and use the Iwasawa
factorization to represent an element $g\in SL(2,\mathbb{R})$ in the
parameterized form
\begin{equation}
  T(g(y,z,\theta)) = \rme^{y \hat{S}_3} \rme^{z \hat{S}_-} \rme^{\rmi
  \theta \hat{S}_2} \, ,
\end{equation}
where $\hat{S}_- = \hat{S}_1 - \rmi \hat{S}_2$.  Let $|0\rangle \in
\mathcal{L}^2(SU(2))$ be a normalized state such that
\begin{equation}
  \langle 0| \hat{S}_3|0\rangle = M \, ,\qquad 
  \langle 0| \hat{S}_1|0\rangle = \langle 0| \hat{S}_2|0\rangle = 0\,,
  \label{eq:ClassicalSU2State}
\end{equation}
where $M$ is real.  We then obtain the classical realization of the
$su(2)$ algebra, $\hat{S}_i \to \mathcal{S}_i$, 
as functions on a cylinder
\begin{equation}
  \label{eq:ClassicalSU2}
  \eqalign{
  \mathcal{S}_1(z,\theta) = \langle 0|\rme^{z\hat{S}_-} \rme^{\rmi\theta
    \hat{S}_2} \hat{S}_1 \rme^{-\rmi\theta \hat{S}_2} 
  \rme^{-z\hat{S}_- }
    |0\rangle = M(\sin\theta - z\cos\theta) \, , \\
  \mathcal{S}_2(z,\theta) = \langle 0|\rme^{z\hat{S}_-} \rme^{\rmi\theta
    \hat{S}_2} \hat{S}_2 \rme^{-\rmi\theta\hat{S}_2} 
  \rme^{-z\hat{S}_-} |0\rangle = \rmi Mz \, , \\
  \mathcal{S}_3(z,\theta) = \langle 0|\rme^{z\hat{S}_-}
    \rme^{\rmi\theta\hat{S}_2} \hat{S}_3 \rme^{-\rmi\theta\hat{S}_2}
    \rme^{-z\hat{S}_- } |0\rangle = M(\cos\theta + z \sin\theta) \,. }
\end{equation}
The Poisson bracket of these functions 
\begin{equation} 
  \{ \mathcal{S}_i,\mathcal{S}_j\}(z,\theta)
  = -\rmi \langle 0| \rme^{z\hat{S}_-}
    \rme^{\rmi\theta\hat{S}_2} [\hat{S}_i,\hat{S}_j]
  \rme^{-\rmi\theta\hat{S}_2} \rme^{-z\hat{S}_-} |0\rangle \,,
\end{equation}
gives
 \begin{equation}
  \{ \mathcal{S}_i,\mathcal{S}_j\} = \mathcal{S}_k \,,\quad i,j,k\;
  {\rm cyclic}.
\end{equation}
It can also be expressed in the classical form
\begin{equation}
  \{ \mathcal{S}_i,\mathcal{S}_j\}(z,\theta)
  = \frac{\rmi}{M} \left( \frac{\partial \mathcal{S}_i}{\partial z}
  \frac{\partial \mathcal{S}_j}{\partial\theta}- \frac{\partial
  \mathcal{S}_i}{\partial\theta} \frac{\partial \mathcal{S}_j}
  {\partial z}\right) . 
\end{equation}

The quantizability condition is that $2M$ should be an integer.
Prequantization is then given by choosing $|\varphi\rangle$ to be an
eigenstate of $\hat S_3$ with eigenvalue $M$ (a half integer) so that
\begin{equation}
  \langle\varphi| \rme^{\rmi\sigma \hat S_3} \rme^{z\hat S_-}
    \rme^{\rmi\theta \hat S_2 }
  |\psi\rangle = \rme^{\rmi M\sigma}\langle\varphi| \rme^{z\hat S_-}
  \rme^{\rmi\theta \hat S_2 } |\psi\rangle \, .
\end{equation}
Coherent state wave functions for the induced representation are
now defined by
\begin{equation}
  \psi (z,\theta) = \langle\varphi| \rme^{z\hat S_-} \rme^{\rmi\theta
    \hat S_2 } |\psi\rangle \, ,\quad |\psi\rangle \in
    \mathcal{L}^2(SU(2)) \, ,
\end{equation}
and the corresponding representation of the infinitesimal generators of
$SU(2)$ is defined in the usual way by
\begin{equation}
  [\Gamma (S_i)\psi] (z,\theta) = \langle \varphi| \rme^{z\hat S_-}
  \rme^{\rmi\theta \hat S_2} \hat S_i|\psi\rangle \, .
\end{equation}
This equation gives $\Gamma(S_2)$ immediately as
\begin{equation}
  \Gamma ({S_2}) = -\rmi \frac{\partial}{\partial \theta} \, .
\end{equation}
From
\begin{eqnarray}
  {[}\Gamma ({S_1})\psi] (z,\theta) &= \langle\varphi| \rme^{z\hat
  S_-}\rme^{\rmi\theta \hat S_2} \hat S_1 |\psi\rangle \nonumber \\
  &= \langle\varphi| \rme^{z\hat S_-}
  [\hat S_1\cos\theta  + \hat S_3 \sin\theta ] \rme^{\rmi\theta \hat S_2}
  |\psi\rangle \, , \\
  {[}\Gamma ({S_3})\psi] (z,\theta) &=\langle\varphi| \rme^{z\hat S_-}
  [\hat S_3\cos\theta  - \hat S_1 \sin\theta ] \rme^{\rmi\theta \hat S_2}
  |\psi\rangle \, ,
\end{eqnarray} 
and the observation that
\begin{equation}
  \eqalign{ 
  \rme^{z\hat S_-} \hat S_1 = \rme^{z\hat S_-} (\hat S_- +\rmi\hat
  S_2) \, ,\\ 
  \rme^{z\hat S_-} \hat S_3 = [\hat S_3 + z \hat S_-] \rme^{z\hat S_-}
  \, ,} 
\end{equation}
it follows that
\begin{equation}
  \eqalign{
  \Gamma ({S_1}) = \sin \theta \big[ M + z\frac{\partial}{\partial
   z}\big] +
  \cos \theta\big[\frac{\partial}{\partial z} + \frac{\partial}{\partial
  \theta}\big] \, , \\
  \Gamma ({S_3}) = \cos \theta \big[ M + z\frac{\partial}{\partial
   z}\big] -
  \sin \theta\big[\frac{\partial}{\partial z} + \frac{\partial}{\partial
  \theta}\big] \, .}
\end{equation}

An irreducible subrepresentation results if $|\varphi\rangle$ is
chosen to be a highest weight state, so that $2M$ is a positive
integer (which we now call $2S$), and satisfies the equations
\begin{equation} 
  \hat S_3 |\varphi\rangle = S |\varphi\rangle \, ,\qquad
  \hat S_+ |\varphi\rangle = 0 \, , 
  \label{eq:SU2highestweight}
\end{equation} 
where
\begin{equation}   
  \hat S_+ = \hat S_1 + \rmi \hat S_2
\end{equation} 
is the adjoint of $\hat S_-$.  The coherent state wave functions then
become independent of $z$,
\begin{equation} 
  \psi(z,\theta) =\langle \varphi|  \rme^{z\hat S_-}
  \rme^{\rmi\theta\hat S_2} |\psi \rangle = \langle\varphi|
  \rme^{\rmi\theta\hat S_2} |\psi \rangle \, ,
\end{equation}
and are seen to be functions on the circle.  The coherent state
representation reduces to
\begin{equation}
  \eqalign{
  \Gamma (\hat S_2 ) = -\rmi \frac{\partial}{\partial \theta} \, , \\
  \Gamma (\hat S_1 ) =  S \sin \theta + \cos \theta
  \frac{\partial}{\partial \theta} \, , \\
  \Gamma (\hat S_3 ) =  S \cos \theta - \sin \theta
  \frac{\partial}{\partial \theta} \, ,}
\end{equation}
which is that of an $su(2)$ irrep of angular momentum $S$.  It
corresponds to the quantization obtained by choosing the polarization
$\mathfrak{p}$ to be the Borel subalgebra of $su(2)^c$ spanned by
$S_3$ and $S_-$.

The inner product and Hilbert space for this irrep are inferred in
coherent state theory from the requirement that, for a unitary
representation, $\Gamma(S_+)$ should be the Hermitian adjoint of
$\Gamma(S_-)$ and vice versa.  Thus, an orthonormal basis
$\mathcal{H}$ is constructed by the systematic methods of K-matrix
theory~\cite{rowe88}.  The inner product is also given in integral
form by the methods of~\cite{triplets}.

\subsubsection{Representation by holomorphic functions}

Equivalent holomorphic representations are obtained by choosing the
same polarization but a different factorization of an
$SL(2,\mathbb{R})$ group element
\begin{equation}
  T(g(x,y,z)) = \rme^{x \hat{S}_3} \rme^{y \hat{S}_-} \rme^{z
  \hat{S}_+} \, .
\end{equation}
Then, for $|0\rangle$ again such that
equation~(\ref{eq:ClassicalSU2State}) is satisfied, we obtain a
classical realization of the $su(2)$ algebra, $\hat{S_i}\to
\mathcal{S}_i$, with
\begin{equation}
  \eqalign{
  \mathcal{S}_1(y,z) = \langle 0|\rme^{y \hat S_-} \rme^{z \hat S_+}
  \, \hat S_1 \, \rme^{-z \hat S_+} \rme^{-y\hat S_-} |0\rangle =
  M(z-y+yz^2) \, ,\\
  \mathcal{S}_2(y,z) = \langle 0|\rme^{y \hat S_-} \rme^{z \hat S_+}
  \, \hat S_2 \, \rme^{-z \hat S_+} \rme^{-y\hat S_-} |0\rangle
  = \rmi M(z+y+yz^2) \, , \\
  \mathcal{S}_3(y,z) = \langle 0|\rme^{y \hat S_-} \rme^{z \hat S_+} \,
  \hat S_3 \, \rme^{-z \hat S_+} \rme^{-y\hat S_-} |0\rangle = M(1+2yz) \, .}
\end{equation}
The Poisson bracket of these functions 
\begin{equation} 
  \{ \mathcal{S}_i,\mathcal{S}_j\}(y,z)
  = -\rmi \langle 0| \rme^{y\hat{S}_-}
    \rme^{z\hat{S}_+} [\hat{S}_i,\hat{S}_j]  \rme^{-z\hat{S}_+}
    \rme^{ -y\hat{S}_-} |0\rangle \,,
\end{equation}
again gives
\begin{equation}
  \{ \mathcal{S}_i,\mathcal{S}_j\} = \mathcal{S}_k \,,\quad i,j,k\;
  {\rm cyclic}\,,
\end{equation}
and is now expressed in the classical form
\begin{equation}
  \{ \mathcal{S}_i,\mathcal{S}_j\}(z,\theta)
  = \frac{\rmi}{2M} \left(
  \frac{\partial \mathcal{S}_i}{\partial y} \frac{\partial
  \mathcal{S}_j}{\partial z} - \frac{\partial \mathcal{S}_i}{\partial
  z} \frac{\partial\mathcal{S}_j}{\partial y}\right) . 
\end{equation}

With $|\varphi\rangle$ an eigenstate of $\hat S_3$ of eigenvalue $M$ (with
$2M$ an integer), coherent state wave functions are given by
\begin{equation}
  \psi (y,z) = \langle\varphi| \rme^{y \hat S_-} \rme^{z \hat S_+}
  |\psi\rangle \, ,\quad |\psi\rangle \in \mathcal{L}^2(SU(2)) \, , 
\end{equation}
and the corresponding representation of the infinitesimal generators of
$SU(2)$ is defined in the usual way by
\begin{equation}
  [\Gamma (S_i)\psi] (y,z) = \langle\varphi| \rme^{y \hat S_-} \rme^{z
  \hat S_+} \hat S_i|\psi\rangle \, .
\end{equation}
One finds that
\begin{equation}
  \eqalign{
  \Gamma (S_+) = \frac{\partial}{\partial z} \, , \\
  \Gamma (S_-) = \frac{\partial}{\partial y}
    +z\Big( 2M+ 2y \frac{\partial}{\partial y}-
  z\frac{\partial}{\partial z}\Big) \, ,\\
  \Gamma (S_3) =  M+  y\frac{\partial}{\partial y}
  - z\frac{\partial}{\partial z} \, .}
\end{equation}
This representation is also obtained by prequantization of the above
classical realization.

An irreducible subrepresentation is again obtained by requiring
$|\varphi\rangle$ to be a highest weight state satisfying
equation~(\ref{eq:SU2highestweight}). The coherent state wave
functions then become independent of $y$,
\begin{equation} 
  \psi(z) =\langle\varphi|  \rme^{y\hat S_-} \rme^{z\hat S_+} |\psi \rangle 
  = \langle\varphi| \rme^{z \hat S_+} |\psi \rangle \, ,
\end{equation}
and holomorphic functions of the variable $z$.  The coherent state
representation reduces to
\begin{equation}
  \Gamma (S_+) = \frac{\partial}{\partial z} \, , \qquad
  \Gamma (S_-) = z\Big( 2S- z \frac{\partial}{\partial
     z}\Big) \, , \qquad
  \Gamma (S_3) =  S - z\frac{\partial}{\partial z}\, .
\end{equation}
which is that of an $su(2)$ irrep of angular momentum $S$. It
corresponds to the quantization obtained by choosing the polarization
$\mathfrak{p}$ to be the Borel subalgebra of $su(2)^c$ spanned by
$S_3$ and $S_-$.

Note that the last two examples involve the same polarization, but
give different realizations.

\subsection{The semidirect product Euclidean group in two dimensions}

The Euclidean group in two dimensions $E(2)\sim [\mathbb{R}^2] SO(2)$
can be realized as the group of translations and rotations in a real
two-dimensional Euclidean space.  Its infinitesimal generators are
two components $(p_x,p_y)$ of a momentum vector and an angular
momentum $L$.  Alternatively, it can be realized as the dynamical
group of a two-dimensional rotor, e.g., a particle moving in a
circle.  A set of observables for such a rotor is given by a pair of
$(x,y)$ coordinate functions and again an angular momentum.  Let $T$
be the regular E(2) representation, with observables in the algebra
satisfing the commutation relations
\begin{equation}
  [\hat{x},\hat{y}]=0 \, ,\qquad 
  [\hat{L},\hat{x}]= \rmi\hbar \hat{y} \, , \qquad 
  [\hat{L},\hat{y}] = -\rmi\hbar \hat{x} \, .
\end{equation}

Group elements in $E(2)$ can be parameterized
\begin{equation}
  T(g(\alpha,\beta,\theta)) = \rme^{-\frac{\rmi}{\hbar} (\alpha\hat
  x+\beta\hat y)} \rme^{-\frac{\rmi}{\hbar}\theta\hat L} \, .
\end{equation} 
Let $|0\rangle$ be a state in the Hilbert space of $T$ having
expectation values
\begin{equation}
  \langle 0| \hat{x}|0\rangle = 0 \, ,\qquad 
  \langle 0| \hat{y}|0\rangle = r \, ,\qquad 
  \langle 0| \hat{L}|0\rangle = 0 \, . 
\end{equation} 
This state defines a classical realization of the $E(2)$ Lie algebra
in which $(x,y,L)$ map to $\beta$-independent functions
$(\mathcal{X}, \mathcal{Y}, \mathcal{L})$ over $g(\alpha, \beta,
\theta)$ defined by
\begin{equation}
  \eqalign{
  \mathcal{X}(\alpha,\theta) = \langle 0| \rme^{-\frac{\rmi}{\hbar}  
  (\alpha\hat x+\beta\hat y)}\rme^{-\frac{\rmi}{\hbar}\theta\hat L}
  \,\hat{x}\, \rme^{\frac{\rmi}{\hbar}\theta\hat L}\rme^{\frac{\rmi}{\hbar}  
  (\alpha\hat x+\beta\hat y)}|0\rangle = r\sin\theta \, , \\
  \mathcal{Y}(\alpha,\theta) = \langle 0|\rme^{-\frac{\rmi}{\hbar}  
  (\alpha\hat x+\beta\hat y)}\rme^{-\frac{\rmi}{\hbar}\theta\hat L}
  \,\hat{y}\, \rme^{\frac{\rmi}{\hbar}\theta\hat L}\rme^{\frac{\rmi}{\hbar}  
  (\alpha\hat x+\beta\hat y)}|0\rangle = r\cos\theta \, , \\
  \mathcal{L}(\alpha,\theta) = \langle 0| \rme^{-\frac{\rmi}{\hbar}  
  (\alpha\hat x +\beta\hat y)}\rme^{-\frac{\rmi}{\hbar}\theta\hat L}
  \,\hat{L}\, \rme^{\frac{\rmi}{\hbar}\theta\hat L}\rme^{\frac{\rmi}{\hbar}  
  (\alpha\hat x+\beta\hat y)}|0\rangle = - \alpha r \, ,}
\end{equation}
and for which the Poisson bracket is given by
\begin{equation}
  \{ \mathcal{A},\mathcal{B}\}(\alpha,\theta) =\frac{1}{r} \Big(
  \frac{\partial \mathcal{A}}{\partial \theta}
  \frac{\partial \mathcal{B}}{\partial \alpha} 
  - \frac{\partial \mathcal{A}}{\partial \alpha} 
  \frac{\partial \mathcal{B}}{\partial \theta}  \Big) \, .
\end{equation} 

Now let $\langle\varphi|$ be a functional on a dense subspace
$\mathbb{H}_D$ of the Hilbert space for the representation $T$ 
such that
\begin{equation}
  \langle\varphi| T(g(\alpha,\beta,\theta))|\psi\rangle =
  \rme^{-\frac{\rmi}{\hbar}\beta r} \langle\varphi|
  \rme^{-\frac{\rmi}{\hbar}\alpha\hat x}\rme^{-\frac{\rmi}{\hbar}\theta
  \hat L}|\psi\rangle \, .
\end{equation}
The space of coherent state wave functions, defined for each
$|\psi\rangle\in\mathbb{H}_D$ by
\begin{equation}
  \psi(\alpha,\theta) = \langle\varphi|\rme^{-\frac{\rmi}{\hbar}
  \alpha\hat x} \rme^{-\frac{\rmi}{\hbar}\theta\hat L}|\psi\rangle \, ,
\end{equation}
is then isomorphic to the space of square integrable functions on a
cylinder with respect to the standard $\rmd\alpha\, \rmd\theta$
measure.  This space carries a reducible representation of $E(2)$ for
which
\begin{equation}
  \eqalign{
  \Gamma(x) = r\sin\theta +\rmi\hbar \cos\theta
  \frac{\partial}{\partial \alpha} \, , \\
  \Gamma(y) = r\cos\theta - \rmi\hbar \sin\theta
  \frac{\partial}{\partial \alpha} \, , \\
  \Gamma(L) = \rmi\hbar \frac{\partial}{\partial\theta} \, .}
\end{equation}
This representation is the same as that obtained for $E(2)$ by
prequantization of the rotor.

To obtain an irreducible representation, one must choose a subalgebra
that in geometric quantization defines a polarization.  A suitable
subalgebra is the Lie algebra of the normal subgroup
$\mathbb{R}^2\subset E(2)$.  Let $\langle\varphi|$ be a functional on
a dense subspace of the Hilbert space of $T$ such that
\begin{equation}
  \langle\varphi| T(\alpha,\beta,\theta)|\psi\rangle =
  \rme^{-\frac{\rmi}{\hbar}r\beta} \langle\varphi| 
  \rme^{-\frac{\rmi}{\hbar}\theta\hat L}|\psi\rangle \, .
\end{equation}
The space of coherent state wave functions, defined for each
$|\psi\rangle\in\mathbb{H}_D$ by
\begin{equation}
  \psi(\theta) = \langle\varphi|\rme^{-\frac{\rmi}{\hbar}\theta\hat
  L}|\psi\rangle \, ,
\end{equation}
is now $\mathcal{L}^2(SO(2))$, the space of square integrable
functions on the circle with respect to the standard $\rmd\theta$
measure.  The coherent state representation of the $E(2)$ Lie algebra
is now irreducible on $\mathcal{L}^2(SO(2))$ and given by
\begin{equation}
  \Gamma(x) = r\sin\theta \, ,\qquad 
  \Gamma(y) = r\cos\theta \, ,\qquad 
  \Gamma(L) = \rmi\hbar\frac{\partial}{\partial\theta} \, .
\end{equation}

\section{Concluding remarks} 

The theory of geometric quantization provides a sophisticated
perspective on the underlying principles for quantization of a
classical model.  On the other hand, the theory of induced
representations is one of the most versatile procedures for
constructing representations of Lie groups and Lie algebras.  As
emphasized by researchers in both fields, the two theories have much
in common and both contribute substantially to the description of
quantum systems.  Unfortunately, because of their formidable
mathematical expressions, they are not readily accessible to most
physicists. Thus, it is useful to know that both theories can be
expressed in the language of coherent state representations, a
language that has been specifically developed to provide practical
methods for performing algebraic calculations in physics.

It has been shown in this paper that the coherent state construction
yields the three types of representations of the SGA of an algebraic
model involved in a quantization scheme: classical realizations,
prequantizations, and full quantizations (unitary irreps).  Examples
have also been given to illustrate how the coherent state approach
provides an intuitive path through the techniques of geometric
quantization for algebraic models.  Thus, we are optimistic that the
coherent state methods presented here will serve to make the methods
of induced representations and geometric quantization accessible to a
wider community.  By expressing the methods of geometric quantization
in the language of coherent state theory, we are also optimistic that
the many techniques developed for the practical application of induced
representation theory to the solution of physical problems will be
equally useful for practical applications of the methods of geometric
quantization.

It is interesting that the classical representations given by coherent
state methods automatically take into account the inherent
limitations, imposed by the uncertainty principle, on an
experimentalist's ability to measure an observable precisely.

In the conventional interpretation of quantum mechanics, the
expectation $\mathcal{X}=\langle\psi|\hat X|\psi\rangle$ of an
observable is identified with the mean value of many (precise)
measurements of the value of the observable when the system is in
a state $|\psi\rangle$. Thus, the
distribution of experimental values,  using an
ideal measurement in which the only limitations on accuracy are
quantum mechanical, is given by the variance
\begin{equation}
  \sigma^2(\mathcal{X}) = \langle \psi|(\hat X 
  - \mathcal{X})^2|\psi\rangle \,.
\end{equation}
It is remarkable then that the mean expectation values of the
observables of a SGA define functions on a coadjoint orbit which are
precisely those of a classical representation.

It is also noted that a given unitary irrep of a SGA can give rise to
many classical realizations.  For example, if a freely rotating object
had squared angular momentum given by ${L(L+1)}$ in a unitary SO(3)
irrep, then the corresponding classical value given by $\sum_k
\mathcal{L}_k^2$ with ${\cal L}_k=\langle \psi |\hat L_k|\psi\rangle$
can have values ranging from zero to $L^2$.  The maximum value of
$L^2$ would be obtained when $|\psi\rangle$ is a minimum uncertainty
(e.g., a highest weight) state.  If an experimentalist could put the
object into such a minimal uncertainty state, then he/she would obtain
an integer value for $L$ and have determined the quantal state of the
rotor precisely.  However, in a practical situation, the uncertainties
in a given experimental situation inevitably exceed the minimal
uncertainties permitted by quantum mechanics.

It is now known that scalar coherent theory has a natural
generalization to a vector coherent theory~\cite{vcs} in which an
irrep of a Lie algebra is induced from a multidimensional irrep of a
subalgebra.  In a sequel to this paper we shall show that VCS theory
can also be expressed in the language of geometric quantization and
that it corresponds to the quantization of a model with intrinsic
degrees of freedom.

\ack

We thank C Bahri for useful discussions.  We are also indebted to G
Rosensteel and the referees for bringing important references to our
attention.  SDB acknowledges the support of a Macquarie University
Research Fellowship.  This paper was supported in part by NSERC of
Canada.
 
\appendix

\section{The covariant derivative and gauge potential}
\label{sec:AppendixA}

If $\mathcal{A}$ is a classical representation of an element $A\in
\mathfrak{g}$ as a function on $\mathcal{O}_\rho = H_\rho \backslash
G$ then the corresponding Hamiltonian vector field $X_{\mathcal{A}}$
on $\mathcal{O}_\rho$ is defined to operate on a function $f$ on
$\mathcal{O}_\rho$ such that $X_{\mathcal{A}}f$ is equal to the
Poisson bracket $\{ \mathcal{A},f\}$.  This requirement means that
$X_{\mathcal{A}}$ must satisfy
\begin{equation}
  [ X_{\mathcal{A}}f](g) = \sum_{\mu\nu}
  \partial_\mu \mathcal{A}(g)\,
  \omega^{\mu\nu}\, \partial_\nu f(g)
  =\sum_{\nu} A^\nu (g)\,\partial_\nu\mathcal{B}(g) \, ,
\end{equation}
where $A^\nu(g)$ is a coefficient in the expansion
\begin{equation}
  A(g)\equiv {\rm Ad}_g(A) = \sum_i A^i(g) A_i + \sum_\nu A^\nu(g)
  A_\nu 
\end{equation}
and $\partial_\nu f$ is defined by
\begin{equation}
  \partial_\nu f(g) = \frac{\partial}{\partial x^\nu} f\big(
  \rme^{-\frac{\rmi}{\hbar} \sum_\alpha x^\alpha  A_\alpha} g\big)
  \Big|_{x=0}\,.
\end{equation}

\noindent {\bf Claim.} {\it Let $\mathcal{A}$ be a classical
  representation of an element $A\in \mathfrak{g}$ as a function on
  $\mathcal{O}_\rho = H_\rho \backslash G$ and $X_{\mathcal{A}}$ the
  corresponding Hamiltonian vector field on $\mathcal{O}_\rho$.  Then
  the covariant derivative $\nabla_{X_{\mathcal{A}}}$ (cf equation
  (48)) acts on a coherent state wave function by
  \begin{equation} 
    \label{eq:CovDeriv}
    [\nabla_{X_{\mathcal{A}}} \psi](g)  = \sum_\nu A^\nu (g) 
    \Big( \partial_\nu +\frac{\rmi}{\hbar} \rho (A_\nu)
    \Big)\psi(g) .
  \end{equation}
  It is  expressible in the form
  \begin{equation}
    \nabla_{X_{\mathcal{A}}} =X_{\mathcal{A}} + \frac{\rmi}{\hbar}
    \theta(X_{\mathcal{A}}) \,,
  \end{equation}
  where $\theta$ is a symplectic potential (one-form) for
  $\mathcal{O}_\rho$. } \medskip

Before proving this claim, we consider first the expression of the
vector field $X_{\mathcal{A}}$ as a derivation in terms of local
coordinates by means of the following observation.

\medskip 
\noindent{\bf Observation.} {\it If $X(x) = -\frac{\rmi}{\hbar} \sum_\mu
x^\mu A_\mu$, then
\begin{equation}
  \rmi\hbar\, \displaystyle\frac{\partial \rme^{X(x)}}{\partial x^\nu}=
  A_\nu(x) \rme^{X(x)} = \rme^{X(x)}A_\nu(-x)\,,
\label{eq:A.deriv} 
\end{equation}
and
\begin{equation} 
  \label{eq:B.deriv} 
  \rmi\hbar\, \displaystyle\frac{\partial \rme^{-X(x)}}{\partial x^\nu}=-
  A_\nu(-x) \rme^{-X(x)} = -\rme^{-X(x)}A_\nu(x)\,,
\end{equation}
where
\begin{equation}
   A_\nu(x) = A_\nu + \frac{1}{2!} [X(x),A_\nu] + 
   \frac{1}{3!} [X(x),[X(x),A_\nu]]+ \cdots
\end{equation} } 
 
\medskip
\noindent{\bf Proof.} The first identity of equation~(\ref{eq:A.deriv})
follows from the observation that
\begin{equation}
  \frac{\partial}{\partial x^\nu} \Bigl(\rme^{-X(x)} \rme^{X(x)}\Bigr)
  =0 \, ,
\end{equation}
and hence that 
\begin{equation}
  \rmi\hbar\frac{\partial}{\partial x^\nu} \rme^{X(x)} =
  \rmi\hbar\Big(\frac{\partial}{\partial x^\nu} - \rme^{X(x)}
  \frac{\partial}{\partial x^\nu} \rme^{-X(x)}\Big)  \rme^{X(x)}\,.
\end{equation}
The second identity of equation~(\ref{eq:A.deriv}) is obtained directly from
\begin{equation}
  \frac{\partial}{\partial x^\nu}\rme^{X(x)} =
  \rme^{X(x)} \Big(\rme^{-X(x)}\frac{\partial}{\partial
  x^\nu}\rme^{X(x)}\Big) \,.
\end{equation}
Equation~(\ref{eq:B.deriv}) is obtained similarly.  
\hfill \emph{QED} \medskip

\medskip 

If $g(x) = \rme^{X(x)} g$ then it follows from the observation
that to leading order in $\delta x$,
\begin{equation}
  g(x+\delta x) = \exp\Big[ -\frac{i}{\hbar}\sum_\mu \delta x^\nu A_\nu(x)
  \Big] g(x) .
\end{equation}
Hence, with the expansion
\begin{equation} 
  \label{eq:AnuExpansion}
  A_\nu(x) = \sum_\mu \Lambda_\nu^\mu (x) A_\mu + \sum_i
  \lambda_\nu^i (x) A_i \,,
\end{equation}
the observation leads to the expression for the derivatives of a
function $f$ on $G$
\begin{equation} 
  \label{eq:xDeriv}
  \frac{\partial}{\partial x^\nu} f(g(x)) =
  \sum_\mu \Lambda_\nu^\mu(x)\,\partial_\mu f( g(x))\color{black} +
  \sum_i \lambda_\nu^i(x)\,\partial_i f( g(x))\color{black} \, . 
\end{equation}
However, when $f=\mathcal{B}$
(a classical observable in the SGA), the second term on the right is
zero, due to the fact that $\mathcal{B}(g)$ satisfies the equations
$\mathcal{B}(hg) = \mathcal{B}(g) $, and hence
\begin{equation}
  \partial_i\, \mathcal{B}(g) =0\,.
\end{equation}
Thus, we obtain 
\begin{equation}
  \partial_\nu\mathcal{B}(g(x)) =\sum_\mu
  \overline\Lambda_\nu^\mu(x)\frac{\partial}{\partial x^\mu}
  \mathcal{B}(g(x))
\end{equation}
and
\begin{equation}
  \label{eq:XHamLocal}
  [ X_{\mathcal{A}}\mathcal{B}](g(x)) 
  = \sum_{\mu\nu} A^\nu (g(x))
  \overline\Lambda_\nu^\mu(x)\frac{\partial}{\partial x^\mu}
  \mathcal{B}(g(x)) \, ,
\end{equation}
where $\overline\Lambda(x)$ is the inverse of the matrix $\Lambda(x)$.

\medskip\noindent{\bf Proof of the Claim.}  The action of $
X_{\mathcal{A}}$, defined by equation (\ref{eq:XHamLocal}) can be
extended to coherent state wave functions.  However, while the action
(\ref{eq:XHamLocal}) of $ X_{\mathcal{A}}$ on functions over
$\mathcal{O}_\rho \sim H_\rho\backslash G$ is covariant, the action on
coherent state wave functions is not covariant; these wave functions
are not defined on $\mathcal{O}_\rho$ but have extra phase factors
and, as a consequence, they are not gauge invariant.  In particular,
$\partial_i \psi(g)$ is not generally zero.  Thus, in evaluating the
covariant derivative, defined by equation~(\ref{eq:CovDeriv}), both
terms on the right hand side of equation~(\ref{eq:xDeriv}) must be
included to give
\begin{equation}
  \label{eq:145}
  \partial_\nu\psi( g(x)) = \sum_\mu
  \overline\Lambda^\mu_\nu(x)
  \Big(\frac{\partial}{\partial x^\mu} +\frac{\rmi}{\hbar} \sum_i
  \lambda_\mu^i \rho(A_i) \Big) \psi(g(x)) \, .
\end{equation}
The covariant derivative of $\psi$ at $g(x)$ is then
\begin{equation}
  [\nabla_{X_{\mathcal{A}}} \psi](g(x))  =
  \sum_\nu A^\nu (g(x))\overline\Lambda^\mu_\nu(x) 
  \Big(\frac{\partial}{\partial x^\mu} 
  +\frac{\rmi}{\hbar} \theta_\mu(x) \Big) \psi(g(x)) \, ,
\end{equation}
where 
\begin{equation}
  \label{eq:thetanu}
  \theta_\mu(x) = \sum_\nu \Lambda^\nu_\mu(x) \rho(A_\nu) + \sum_i
  \lambda_\mu^i(x) \rho(A_i)= \rho(A_\mu(x))\,. 
\end{equation}
Thus, with the interpretation of $\theta_\nu$ as a component of a
one-form $\theta_{g(x)} = \sum_\nu \theta_\nu(x)\, \rmd x^\nu$, so that
\begin{equation}
  \theta_\nu (x) = \theta_{g(x)} (\partial /\partial x^\nu) \,,
\end{equation}
the covariant derivative is expressed $\nabla_{X_{\mathcal{A}}}
=X_{\mathcal{A}} + \frac{\rmi}{\hbar} \theta(X_{\mathcal{A}})$ as
claimed.

It remains to be shown that the one-form $\theta$ is a symplectic
potential, i.e., that the symplectic form on $\mathcal{O}_\rho$ is
given by
\begin{equation}  
  \label{eq:SympForm}
  \omega_{g(x)} = \rmd\theta_{g(x)}=\sum_{\mu\nu} \frac{\partial
  \theta_\nu(x)}{\partial x^\mu} \, \rmd x^\mu \wedge \rmd x^\nu \,.
\end{equation}

First observe, from equation~(\ref{eq:thetanu}), that
\begin{equation}
  \frac{\partial \theta_\nu(x)}{\partial x^\mu} = 
  \rho \Big( \frac{\partial A_\nu(x)}{\partial x^\mu}\Big) \,.
\end{equation}
Now, with $A_\nu(x)$ written in the form
\begin{equation}
  A_\nu(x) =  - \rme^{X(x)} \rmi\hbar\, \frac{\partial}{\partial
  x^{\nu}} \rme^{-X(x)} \,,
\end{equation}
it follows, from multiple use of the observation, that
\begin{equation}
  \frac{\partial A_\nu(x)}{\partial x^\mu} =
  -\frac{\rmi}{\hbar} [A_\mu(x),A_\nu(x)] \,.
\end{equation}
Thus, we obtain
\begin{equation}
  \frac{\partial \theta_\nu(x)}{\partial x^\mu} = 
  -\frac{\rmi}{\hbar} \rho([A_\mu(x),A_\nu(x)])\,,
\end{equation}
which from the definition of $A_\nu(x)$
(equation~(\ref{eq:AnuExpansion})) and $\omega_{\mu\nu}$
(equation~(\ref{eq:omega})) gives
\begin{equation}
  \frac{\partial \theta_\nu(x)}{\partial x^\mu} = 
  \sum_{\mu',\nu'} \Lambda^{\mu'}_\mu(x) \omega_{\mu',\nu'}
  \Lambda^{\nu'}_\nu (x)\,.
\end{equation}
Therefore, if $\omega$ is the two-form defined by
equation~(\ref{eq:SympForm}), then for the vector fields defined by
equation~(\ref{eq:XHamLocal}), we obtain
\begin{equation}
  \omega_{g(x)} (X_{\mathcal{A}},X_{\mathcal{B}}) = \sum_{\mu\nu}
  A^\mu(g(x))\, \omega_{\mu\nu} B^{\nu}(g(x)) \,,
\end{equation}
which is identical to the expression in equation~(\ref{eq:PBracket})
for the Poisson bracket $\{ \mathcal{A},\mathcal{B}\}(g(x))$.  This
result confirms that the one-form $\theta$ is indeed a symplectic
potential and completes the proof of the claim.  \hfill \emph{QED}


\medskip

\section*{References}

\begin{thebibliography}{99}
  
\bibitem{K1} Kirillov A A 1962 {\em Unitary representations of
    nilpotent Lie groups}, Russian Math.\ Surveys {\bf 17} 57-110;
  {\it Usp.\ Mat.\ Nauk.}\ {\bf 106} 57

\bibitem{kostant} Kostant B 1970 {\it On Certain Unitary
    Representations which Arise from a Quantization Theory}, in {\it
    Group Representations in Mathematics and Physics, Lecture Notes in
    Physics, Vol.\ 6} (Berlin: Springer); {\it Quantization and
    Unitary Representations}, in {\it Letures in Modern Analysis and
    Applications III, Lecture Notes in Mathematics, Vol.\ 170}
  (Berlin, Springer)

\bibitem{souriau} Souriau J--M 1966 {\it Comm.\ Math.\ Phys.}\  
  \textbf{1} 374;  1970 {\it Structure des syst\`emes
  dynamiques} (Paris: Dunod)

\bibitem{thesis} Bartlett S D 2000 {\it Quantization of a Classical Model
  with Symmetry} doctoral thesis, University of Toronto, Toronto,
  Canada

\bibitem{mackey} Mackey G W 1952 {\it Ann.\ of Math.}\ 
  \textbf{55} 101; 1968 {\it Induced Representation of Groups and
  Quantum Mechanics} (New York: Benjamin); 1978 {\it Unitary Group
  Representations in Physics, Probability and Number Theory}
  (Reading, MA: Benjamin)

\bibitem{per86} Perelomov A 1986 {\it Generalized Coherent States and Their
  Applications} (Berlin: Springer-Verlag)

\bibitem{Klauder} Klauder J R and Skagerstam B-S 1985 {\it
Coherent States; Applications  in Physics and Mathematical Physics}
(Singapore: World Scientific)

\bibitem{Onofri} Onofri E 1975 \JMP \textbf{16} 1087

\bibitem{RR91} Rowe D J and Repka J 1991 \JMP \textbf{32} 2614

\bibitem{OnofriPauri} Onofri E and Pauri M 1972 \JMP
  {\bf 13} 533; {\it Lett.\ Nuovov Cim.}\ {\bf 2}(3) 35

\bibitem{Wig39} Wigner E P 1939 {\it Ann.\ Math.}\ {\bf 40} 149

\bibitem{Zak} Zak J 1960 \JMP {\bf 1} 165
  
\bibitem{diffeo} Goldin G A, Menikoff R and Sharp D H 1987 \PRL
  \textbf{58} 2162; Goldin G A 1992 {\it Internat. J. Modern Phys.
    B}\textbf{6} 1905; Penna V and Spera M 1989 \JMP \textbf{30} 2778;
  Penna V and Spera M 1991 \JMP \textbf{33} 901
  
\bibitem{RI79} Rosensteel G and Ihrig E 1979 {\it Ann.\ Phys.\ (N.Y.)}
  {\bf 121} 113.

\bibitem{zhao} Zhao Q 1998 {\it Commun.\ Math.\ Phys.}\
  \textbf{194} 135

\bibitem{GS}  Guillemin V and Sternberg S 1984 {\it Symplectic Techniques in
   Physics} (Cambridge: Cambridge University Press)

\bibitem{Glauber} Glauber R Y 1963 \PR {\bf 131} 2766

\bibitem{vcs} Rowe D J 1984 \JMP \textbf{25} 2662;  Rowe D J,
  Rosensteel G and Carr R 1984 \JPA \textbf{17} L399; Rowe D J,
  Wybourne B G and Butler P H 1985 \JPA \textbf{18} 939; Rowe D J,
  Rosensteel G and Gilmore R 1985 \JMP \textbf{26} 2787

\bibitem{Rawnsley} Rawnsley W 1977 {\it Quart.\ J.\ Math., Oxford}\ 
  {\bf 28} 403; Rawnsley J, Cahen M and Gutt S 1990 {\it J. Geom.
    Phys.}\ \textbf{7} 45

\bibitem{kirillov} Kirillov A A 1975 {\it Elements of the Theory
  of Representations} (Berlin: Springer/Verlag)

\bibitem{Kir99} Kirillov A 1999 {\it Bull.\ Am.\ Math.\ Soc.}\
  \textbf{36} 433

\bibitem{streater} Streater R F 1967 {\it Comm. Math. Phys.}\
  \textbf{4} 217

\bibitem{dunne} Dunne S A 1969 \JMP \textbf{10} 860

\bibitem{joseph}
  Joseph A 1970 {\it Commun.\ Math.\ Phys.}\ \textbf{17} 210;
  Gotay M J 1995 {\it Quantization, coherent states, and complex
  structures (Bialowieza 1994)} (New York: Plenum);
  Gotay M J, Grundling H B and Hurst C A 1996 
  {\it Trans.\ Amer.\ Math.\ Soc.}\ \textbf{348} 1579;
  Gotay M J and Grundling H B 1997 {\it Rep.\ Math.\ Phys.}\
  \textbf{40} 107

\bibitem{ihrig} Ihrig E and Rosensteel G 1993 {\it Int.\
  J.\ Theoret.\ Phys.} \textbf{32} 843 

\bibitem{marsden} Marsden J E and Ratiu T S 1994 {\it Introduction to
    Mechanics and Symmetry} (New York:  Springer-Verlag)

\bibitem{diractext} Dirac P A M 1958 {\it The Principles of
  Quantum Mechanics} (Oxford: Oxford University Press)

\bibitem{woodhouse} Woodhouse N M J 1991 {\it Geometric
  Quantization} (Oxford: Oxford University Press)

\bibitem{kashiwara} Kashiwara M and Vergne M 1978
  {\it Invent.\ Math.}\ \textbf{44} 1

\bibitem{HarmAnalysis}  Wallach N 1973 {\it Hamonic analysis on
  homogenious spaces} (New York: Maral Dekker) p 115.

\bibitem{Harish-Chandra} Harish-Chandra 1955 {\it Amer.\ J.\ Math.}\ {\bf 
77} 743; 1956 {\bf 78} 1; 1956 {\bf 78} 564

\bibitem{auslander} Auslander L and Kostant B 1971 {\it
    Invent. Math.}\ \textbf{14} 255

\bibitem{rowe88} Rowe D J, Le Blanc R and Hecht K T 1998 \JMP
  \textbf{29} 287; Rowe D J 1995 \JMP \textbf{36} 1520 

\bibitem{triplets} Rowe D J and Repka J in preparation

\bibitem{BS} Bargmann V 1961 {\it Commun.\ Pure and Appl.\ Math.}\ {\bf 14}
   187 ; Segal I 1963 {\it Mathematical Problems of Relativistic
   Physics} (Amer. Math. Soc. Publ., Providence,  Rhode Island)

\end{thebibliography}
\end{document}